\documentclass[aps,
physrev,
superscriptaddress,
amsmath,
amssymb,
reprint,
nofootinbib, 
floatfix]
{revtex4-2}

\usepackage[utf8]{inputenc}
\usepackage{graphicx}
\usepackage{color}
\usepackage{comment}
\usepackage{url}
\usepackage[breaklinks=true]{hyperref}
\usepackage[italicdiff]{physics}
\usepackage[version=4]{mhchem}

\newcommand{\rp}{{r^\prime}}
\newcommand{\mr}[1]{\mathrm{#1}}
\newcommand{\mc}[1]{\mathcal{#1}}

\date{\today}

\begin{document}

\title{
   Quantum expectation-value estimation by computational basis sampling
}

\author{Masaya Kohda}
\email{kohda@qunasys.com}
\affiliation{QunaSys Inc., Aqua Hakusan Building 9F, 1-13-7 Hakusan, Bunkyo, Tokyo 113-0001, Japan}

\author{Ryosuke Imai}
\affiliation{QunaSys Inc., Aqua Hakusan Building 9F, 1-13-7 Hakusan, Bunkyo, Tokyo 113-0001, Japan}

\author{Keita Kanno}
\affiliation{QunaSys Inc., Aqua Hakusan Building 9F, 1-13-7 Hakusan, Bunkyo, Tokyo 113-0001, Japan}

\author{Kosuke Mitarai}
\affiliation{Graduate School of Engineering Science, Osaka University, 1-3 Machikaneyama, Toyonaka, Osaka 560-8531, Japan}
\affiliation{
Center for Quantum Information and Quantum Biology, Osaka University, 1-2 Machikaneyama, Toyonaka, Osaka, 560-0043, Japan
}
\affiliation{JST, PRESTO, 4-1-8 Honcho, Kawaguchi, Saitama 332-0012, Japan}

\author{\\Wataru Mizukami}
\affiliation{Graduate School of Engineering Science, Osaka University, 1-3 Machikaneyama, Toyonaka, Osaka 560-8531, Japan}
\affiliation{
Center for Quantum Information and Quantum Biology, Osaka University, 1-2 Machikaneyama, Toyonaka, Osaka, 560-0043, Japan
}
\affiliation{JST, PRESTO, 4-1-8 Honcho, Kawaguchi, Saitama 332-0012, Japan}

\author{Yuya O. Nakagawa}
\email{nakagawa@qunasys.com}
\affiliation{QunaSys Inc., Aqua Hakusan Building 9F, 1-13-7 Hakusan, Bunkyo, Tokyo 113-0001, Japan}

\begin{abstract}
Measuring expectation values of observables is an essential ingredient in variational quantum algorithms.
A practical obstacle is the necessity of a large number of measurements for statistical convergence to meet requirements of precision, such as chemical accuracy in the application to quantum chemistry computations.
Here we propose an algorithm to estimate the expectation value based on its approximate expression as a weighted sum of classically-tractable matrix elements with some modulation, where the weight and modulation factors are evaluated by sampling appropriately prepared quantum states in the computational basis on quantum computers.
Each of those states is prepared by applying a unitary transformation consisting of at most $N$ CNOT gates, where $N$ is the number of qubits, to a target quantum state whose expectation value is evaluated. 
Our algorithm is expected to require fewer measurements than conventional methods for a required statistical precision of the expectation value when the target quantum state is {\it concentrated} in particular computational basis states.
We provide numerical comparisons of our method with existing ones for measuring electronic ground state energies (expectation values of electronic Hamiltonians for the lowest-energy states) of various small molecules.
Numerical results show that our method can reduce the numbers of measurements to obtain the ground state energies for a targeted precision by several orders of magnitudes for molecules whose ground states are concentrated.
Our results provide another route to measure expectation values of observables, which could accelerate the variational quantum algorithms.  
\end{abstract}

\maketitle

\section{Introduction}
With the advent of noisy intermediate-scale quantum (NISQ) devices~\cite{preskill2018quantum}, there are intense studies on variational quantum algorithms (see, e.g., \cite{cerezo2021,endo2021hybrid,tilly2021variational} for review).
In particular, the variational quantum eigensolver (VQE)~\cite{peruzzo2014variational} is a promising algorithm in application to condensed matter physics and quantum chemistry~\cite{mcardle2018quantum, cao2019quantum}.
Those hybrid quantum-classical algorithms typically utilize quantum computers to measure expectation values of observables for a given state realized by a parametrized quantum circuit.
A practical obstacle is the necessity to repeat measurements many times to suppress the statistical fluctuation of the expectation value down to a sufficient precision for practical use.
For instance, quantum chemical calculations require an accuracy of $1$ kcal/mol $\simeq 1.6 \times 10^{-3}$~Hartree, the so-called chemical accuracy, to discuss energy differences.
Such a demanding requirement could make VQE impractical due to a huge amount of time needed even for small molecules~\cite{gonthier2020identifying}. 
Thus, it is highly desirable to find a way to measure expectation values of observables with better statistical convergence.

A prototype way~\cite{peruzzo2014variational} to estimate the expectation value $\ev{O}{\psi}$ of an observable $O$ for a quantum state $\ket{\psi}$ on NISQ devices is as follows: first, we expand $O$ in terms of Pauli strings $P_i$, i.e., tensor products of Pauli operators on single qubits; we then measure the expectation value $\ev{P_i}{\psi}$ term by term and those results are assembled by classical post-processing to yield the entire expectation value $\ev{O}{\psi}$.
Estimating the expectation value of each Pauli string with a precision $\epsilon$ requires $\mathcal{O}(1/\epsilon^2)$ measurements, and hence the total number of measurements is $\mathcal{O}(M/\epsilon^2)$, where $M$ is the number of the Pauli strings in the expansion of $O$ (note that the corresponding precision of $\ev{O}{\psi}$ is $\mathcal{O}(\sqrt{M} \epsilon)$).
When the observable $O$ is a Hamiltonian for electrons in a molecule, $M$ is $\mathcal{O}(N^4)$ with $N$ being the number of qubits.
This factor of $M$, albeit polynomial in $N$, can be numerically large, which would be problematic in practical applications to quantum chemistry.
There are various studies to reduce the numbers of measurements to obtain expectation values of observables with a fixed accuracy by, e.g., grouping Pauli strings which can be simultaneously measured~\cite{mcclean2016,kandala2017, jena2019pauli, izmaylov2019,izmaylov2020,verteletskyi2020,yen2020, 
gokhale2020ON3, zhao2020,crawford2021efficient,bonet2020,hamamura2020,huggins2021efficient,
yen2021cartan,shlosberg2021adaptive,yen2022deterministic} 
with or without optimized allocation of measurement budget to each group of Pauli strings~\cite{wecker2015,rubin2018application, arrasmith2020operator}, using 
a technique called classical shadows~\cite{huang2020, zhao2021, hadfield2020measurements, wu2021overlapped}, or decomposing a quantum state into a classically-tractable part and a part dealt with a quantum computer~\cite{radin2021classically}.
However, we still need to greatly reduce the measurement counts for practical applications.

In this study, we propose a hybrid quantum-classical algorithm to efficiently measure the expectation values of observables, focusing on a property of quantum states rather than that of observables.
Our algorithm is designed to be effective when the state is {\it concentrated}, i.e., when only a small number of amplitudes in a measurement basis are non-negligible. 
This relies on an empirical insight that, in many cases, practically-interesting quantum states are concentrated: especially in quantum chemistry problems with second quantization, low-lying eigenstates of a molecular Hamiltonian are often well-described by a small number of computational basis states, or Slater determinants, such as the Hartree-Fock state and excited states proximate to it.
Concretely, by taking the computational basis as the measurement basis, we rewrite the expectation value $\ev{O}{\psi}$ as a weighted sum of transition matrix elements of the observable, modulated by density matrix elements of the state, for the computational basis states (the formula is explicitly given in Eq.~\eqref{eq:ev_O_3}).
While the transition matrix elements can be estimated by classical computers, the weight and modulation factors can be obtained by sampling appropriately-prepared states in the computational basis on quantum computers.
The preparation for each of such states can be performed by applying a unitary transformation consisting of at most $N$ CNOT gates to the state $\ket{\psi}$, where $N$ is the number of qubits.
For a concentrated state, the sum can be approximated with a limited number of computational basis states, which leads to the reduction of the number of quantities to be measured, enabling a faster expectation value estimation.

We also make quantitative comparisons of our method with other methods
by taking electronic Hamiltonians and their ground states of small molecules as examples.
Specifically, we numerically evaluate the variances of estimated expectation values and infer the numbers of measurements to achieve a precision comparable to the chemical accuracy.
We find several illustrative cases where our approach outperforms the conventional methods, reassuring our intuition that certain ground states are concentrated (main results of this work are summarized in Fig.~\ref{fig:shot}).
Our approach focuses on the nature of quantum states rather than observables unlike conventional methods, and hence provides another route to quantum expectation value estimations, which may facilitate efficient executions of variational quantum algorithms.

The rest of this article is organized as follows.
In Sec.~\ref{sec:method}, we first review conventional methods for estimating expectation values of observables and then introduce our algorithm.
In Sec.~\ref{sec:experiment}, numerical comparisons with conventional methods are given by evaluating variances of energy expectation values for small molecules.
In Sec.~\ref{sec:discussion}, we discuss potential advantages of our work and how our method is related to existing classical computational methods.
We summarize this work in Sec.~\ref{sec:summary}.
Formulas for variances of expectation values, details of numerical calculations and discussion on possible improvements of our algorithm are given in Appendices.

\section{Methods
\label{sec:method}}

Our interest in this article is to estimate the expectation values of observables with smaller numbers of measurements.
To be specific, we consider the expectation value of a Hermitian operator $O$ for a state $\ket{\psi}$ defined on $N$ qubits, given by
\begin{align}
\ev{O} = \ev{O}{\psi}.
\label{eq:ev_O_1}
\end{align}
We eventually consider a molecular Hamiltonian and its ground state as the observable and state, respectively, aiming at its application to VQE, but the formalism of our method in this section is not limited to those examples.
We discuss under which condition our method can be effective later in this section.

\subsection{Conventional methods to estimate expectation values
\label{subsec: conventional methods}}

We first recapitulate conventional methods for estimating the expectation values of observables before introducing our method.

The observable $O$ defined on $N$ qubits can be expressed as a linear combination of Pauli strings, or tensor products of Pauli and identity operators on single qubits, $P_i \in \{ I,X,Y,Z \}^{\otimes N}$:
\begin{align}
O = \sum_{i=1}^{M} c_i P_i,
\label{eq:O_pauli}
\end{align}
where $c_i$ are real coefficients and $M$ is the number of the Pauli strings.
The expectation value of $O$ can be written as
\begin{align}
\ev{O}{\psi} = \sum_{i=1}^M c_i \ev{P_i}{\psi}.
\label{eq:ev_O_2}
\end{align}
In the simplest way to estimate the expectation value, each of $\ev{P_i}{\psi}$ is measured on a quantum computer and then assembled into $\ev{O}{\psi}$ by classical post-processing.
If we perform $L_i$ times of measurements for each Pauli string $P_i$, the standard deviation of the estimated $\ev{P_i}{\psi}$ due to a finite $L_i$ is $\epsilon_i^{\mr{naive}} = \sqrt{ (1-\ev{P_i}{\psi}^2) /  L_i}$, as an outcome of the single $P_i$ measurement is either $+1$ or $-1$ and follows the Bernoulli distribution.
The standard deviation of the estimated $\ev{O}{\psi}$ is then given by
\begin{equation}
 \epsilon^{\mr{naive}} = \sqrt{ \sum_{i=1}^M c_i^2 \frac{1 - \ev{P_i}{\psi}^2 }{L_i} }. 
 \label{eq: naive epsilon}
\end{equation}

In this simplest setting, there are $M$ quantities to be measured.
In applications to condensed matter physics and quantum chemistry, for instance, an observable of interest is Hamiltonian.
Such Hamiltonians contain $M = \mathcal{O}(\mr{poly}(N))$ terms of Pauli strings.
This sparsity of the observables is one of the bases for the efficient simulation of quantum systems by quantum computers~\cite{nielsen2010}.
Nevertheless, in the first-principle electronic state calculations of molecules, $M$ is $\mathcal O(N^4)$ and a nominal target of quantum computations would be a simulation of $N\geq100$ that is beyond the capacity of the state-of-the-art classical simulations.
This can result in a numerically large prefactor in the scaling of the standard deviation $\epsilon$ with the total number of measurements, $L=\sum_i L_i$, and would require a practically huge value of $L$ to suppress $\epsilon$ down to the precision comparable to the chemical accuracy.

To reduce the total number of measurements, or equivalently, to suppress the standard deviation $\epsilon$, a plenty of studies have been performed given the estimation of the expectation values is an indispensable subroutine of many variational quantum algorithms~\cite{endo2021hybrid, cerezo2021}.
One strategy is to group Pauli strings which can be simultaneously measured~\cite{mcclean2016,kandala2017, jena2019pauli, izmaylov2019,izmaylov2020,verteletskyi2020,yen2020, 
gokhale2020ON3, zhao2020,crawford2021efficient,bonet2020,hamamura2020,huggins2021efficient,
yen2021cartan,shlosberg2021adaptive,yen2022deterministic}.
The grouping reduces the number of quantities to be measured on quantum computers and can decrease the total number of measurements.
Another strategy is to optimize the value of $L_i$, or the number of measurements performed for $\ev{P_i}{\psi}$, to decrease the overall standard deviation shown in Eq.~\eqref{eq: naive epsilon}~\cite{wecker2015,rubin2018application, arrasmith2020operator,shlosberg2021adaptive}.
By choosing $L_i$ depending on the value of $c_i^2 (1-\ev{P_i}{\psi}^2)$, the overall standard deviation can be minimized with the total number of measurements $L=\sum_i L_i$ fixed.
Note that two strategies, the grouping and the optimization of the number of measurements for each measured quantities, can be employed at the same time.
Despite those efforts, however, further improvements are still awaited for practical applications in quantum chemistry; for instance, a recent work by Gonthier {\it et al.}~\cite{gonthier2020identifying} demonstrates that, even with grouping and optimization techniques, it may take several days to estimate a single energy expectation value even for a small molecule such as methane in a sufficient precision to analyze the combustion energy.
Further technical details of the conventional expectation value estimations can be found in Appendices~\ref{appsubsec: variance conventional} and \ref{appsec: grouping}.

\subsection{Our method
\label{subsec:our-method}}
While the conventional methods are based on the expansion of the observable $O$ by Pauli strings, our method exploits the expansion of the state $\ket{\psi}$ by basis states.
In principle, any basis may be chosen to expand $\ket{\psi}$ as long as (1) transition matrix elements of $O$ for the basis states can be computed classically and (2) single weight factors and interference factors for the basis states, both of which are shortly defined, can be measured on quantum computers.
For illustration, we take the computational basis states $\ket{n}$ ($n = 0, 1, \cdots,2^N-1$) to expand the state $\ket{\psi}$ throughout this article.

\subsubsection{Idea: concentrated states}
By using the computational basis states $\ket{n}$, the state $\ket{\psi}$ can be expressed as
\begin{align}
\ket{\psi}
= \sum_{n=0}^{2^N -1}
   \ip{n}{\psi} \ket{n}.
\label{eq:state_1}
\end{align}
With this expansion, the expectation value may be rewritten as
\begin{align}
\ev{O}{\psi} 
&= \sum_{m,n=0}^{2^N -1}
    \bra{\psi}\ket{m}
    \mel{m}{O}{n}  \bra{n}\ket{\psi} \nonumber
\\
&= {\sum_{m,n}}^{\prime}
    \abs{\bra{m}\ket{\psi}}^2\abs{\bra{n}\ket{\psi}}^2
    \frac{\mel{m}{O}{n} }{\bra{m}\ket{\psi} \bra{\psi}\ket{n}},
    \label{eq:ev_O_3}
\end{align}
where the summation with prime in the last line is meant to exclude terms with $\bra{m}\ket{\psi}=0$ or $\bra{n}\ket{\psi}=0$.
Equation~\eqref{eq:ev_O_3} is central to our proposal though it may look rather innocuous.
We interpret this equation as a weighted sum of $\mel{m}{O}{n} / (\bra{m}\ket{\psi} \bra{\psi}\ket{n})$, or the transition matrix elements of the observable modulated by $\bra{m}\ket{\psi} \bra{\psi}\ket{n}$, with the weight $\abs{\bra{m}\ket{\psi}}^2\abs{\bra{n}\ket{\psi}}^2$.
There are exponentially many terms in the sum, but if only a limited number of the basis states $\ket{n}$ have non-negligible overlaps with the state $\ket{\psi}$, or non-negligible $|\ip{n}{\psi}|^2$, 
the number of terms to be summed up can be significantly reduced for some precision.
In that case, we can truncate the summation in Eq.~\eqref{eq:ev_O_3} keeping only the basis states $\ket{n}$ that have non-negligible $|\ip{n}{\psi}|^2$.
We call a state a {\it concentrated} state in some basis when a small number of the basis states have non-negligible overlaps with the state $\ket{\psi}$ of our interest.
In this explicit demonstration of our idea, we focus on concentrated states in the computational basis.
Our proposal to estimate the expectation value, elaborated in the rest of this section, is effective for concentrated states.

Our observation is that, in many physical and chemical problems of our interest, a target state $\ket{\psi}$ is often concentrated.
Especially in quantum chemistry problems, low-lying electronic eigenstates (including the ground state) of a molecular Hamiltonian are often considered to be described by linear combinations of a small number of Slater determinants, which are translated into the computational basis states on quantum computers under a typical mapping of the molecular Hamiltonian to the qubit Hamiltonian.
In such a case, including only a limited number of the basis states is sufficient to estimate the double sum.
We see this point by concrete numerical examples in Sec.~\ref{sec:experiment}.

There are several ways to calculate the summation in Eq.~\eqref{eq:ev_O_3} by using classical and quantum computers.
One way is to implement the sum a la importance sampling: that is, we draw pairs of numbers $(m, n)$ from a probability distribution $p(m,n)=\abs{\bra{m}\ket{\psi}}^2\abs{\bra{n}\ket{\psi}}^2$ (done by sampling the state $\ket{\psi}$ in the computational basis) and take a sample average of $\mel{m}{O}{n} / (\bra{m}\ket{\psi} \bra{\psi}\ket{n})$ calculated for each pair.
This is reminiscent of the variational Monte Carlo (VMC)~\cite{mcmillan1965,ceperley1977} and we discuss the relationship in Sec.~\ref{sec:discussion}.
Another way is to construct the weight $\abs{\bra{m}\ket{\psi}}^2\abs{\bra{n}\ket{\psi}}^2$ in advance for a small subset of computational basis states and then take the summation with $\mel{m}{O}{n} / (\bra{m}\ket{\psi} \bra{\psi}\ket{n})$ explicitly over the subset.
Both ways can offer efficient calculations of the sum when the state $\ket{\psi}$ is concentrated.
Yet, in this article, we focus on the latter way due to the ease of statistical analysis.

In the latter way to calculate Eq.~\eqref{eq:ev_O_3}, we have to evaluate three elementary quantities by using classical and quantum computers.
The first one is $|\ip{n}{\psi}|^2$, which we call the single weight factor.
The second is the transition matrix element $\mel{m}{O}{n}$.
The third is $\bra{m}\ket{\psi} \bra{\psi}\ket{n}$, which we call the interference factor for a possible complex phase.
The single weight and interference factors can be obtained by quantum computations while the transition matrix elements by classical computations, as explained in the following subsection.

\subsubsection{Evaluation of each component in the summation}

The first quantity, the single weight factor $\abs{\bra{n}\ket{\psi}}^2$, can be estimated by performing the projective measurement on $\ket{\psi}$ in the computational basis.
That is, one repeats a preparation of the state $\ket{\psi}$ on quantum computers followed by the measurement in the computational basis set $\{ \ket{m} \}_{m=0}^{2^N-1}$.
The observed frequency of an outcome ``$n$" gives an estimator of $|\ip{n}{\psi}|^2$.

On the other hand, the second quantity, or the transition matrix element $\mel{m}{O}{n}$, can be obtained solely by classical computation.
We use Eq.~\eqref{eq:O_pauli} to express the transition matrix element as
\begin{align}
\mel{m}{O}{n} 
= \sum_{i=1}^{M} c_i \mel{m}{P_i}{n}.
\end{align}
Because $P_i$ are products of Pauli operators on single qubits and $\ket{m},\ket{n}$ are the computational basis states, $\mel{m}{P_i}{n}$ can be estimated simply by algebra that takes the classical computational time linear in $N$.
The total classical computational time to estimate the matrix element $\mel{m}{O}{n}$ is then $\mathcal{O}(MN)$.
As mentioned before, $M$ is polynomial in $N$ in typical applications;
hence, $\mel{m}{O}{n}$ can be efficiently estimated by classical computation, so long as the observable $O$ is sparse, or contains at most a polynomial number of Pauli strings in the number of qubits $N$.
With the approximation and notation introduced around Eq.~\eqref{eq:state_2}, the total number of $\mel{m}{O}{n}$ to be evaluated is $\mathcal{O}(R^2)$.
In some specific cases, the evaluation of the matrix elements may be performed even faster.
For instance, in the application to quantum chemistry where $O$ is a molecular Hamiltonian and $\ket{m}, \ket{n}$ are Slater determinants, $\mel{m}{O}{n}$ can be more easily evaluated in the fermionic basis directly through the Slater-Condon rules;
in particular, $\mel{m}{O}{n}$ vanishes for the Slater determinants that differ by three or more spin orbitals.

The remaining ingredient is $\bra{m}\ket{\psi} \bra{\psi}\ket{n}$, the interference factor.
When $m=n$, it is equivalent to the single weight factor $\abs{\bra{n}\ket{\psi}}^2$.
When $m\neq n$, it may have a phase and requires a dedicated method to estimate.
Here, we explain one of possible methods for such an estimation.
We first observe that $\bra{m}\ket{\psi}\bra{\psi}\ket{n}$ may be written as
\begin{align}
&\bra{m}\ket{\psi}\bra{\psi}\ket{n}\notag\\
&= \mathcal{A}_{m,n} + i\mathcal{B}_{m,n}
  -\frac{1+i}{2} 
  \left( \abs{\bra{m}\ket{\psi}}^2
  +\abs{\bra{n}\ket{\psi}}^2 \right)
\label{eq:intf_factor_1}
\end{align}
for $m\neq n$, where $\mathcal{A}_{m,n}$ and $\mathcal{B}_{m,n}$ are defined by
\begin{align}
\mathcal{A}_{m,n}
\equiv \abs{\frac{\bra{m} + \bra{n}}{\sqrt{2}} \ket{\psi}}^2,~
\mathcal{B}_{m,n}
\equiv \abs{\frac{\bra{m} + i \bra{n}}{\sqrt{2}} \ket{\psi}}^2.
\end{align}
Let $U_{m,n}$ be a unitary operator such that
\begin{align}
U_{m,n}\pqty{\frac{\ket{m}+\ket{n}}{\sqrt{2}}} = \ket{0},
\end{align}
for a given pair of $m\neq n$.
Then, one may rewrite $\mathcal{A}_{m,n}$ as $\mathcal{A}_{m,n} = \abs{\bra{0}U_{m,n}\ket{\psi}}^2$, implying that $\mathcal{A}_{m,n}$ can be estimated by the probability for obtaining ``0" when measuring the state $U_{m,n}\ket{\psi}$ in the computational basis.
Similarly, $\mathcal{B}_{m,n}$ can be rewritten as $\mathcal{B}_{m,n} = \abs{\bra{0}V_{m,n}\ket{\psi}}^2$ by a unitary operator $V_{m,n}$ satisfying
\begin{align}
V_{m,n}\pqty{\frac{\ket{m}-i\ket{n}}{\sqrt{2}}} = \ket{0},
\end{align}
and can be estimated by the probability for obtaining ``0" when measuring the state $V_{m,n}\ket{\psi}$.
Combining $\mathcal{A}_{m,n}$, $\mathcal{B}_{m,n}$ and the single weight factors $\abs{\bra{m}\ket{\psi}}^2, \abs{\bra{n}\ket{\psi}}^2$, one can estimate the interference factor $\bra{m}\ket{\psi}\bra{\psi}\ket{n}$ for each pair of $m\neq n$ by Eq.~\eqref{eq:intf_factor_1}.

The circuits for $U_{m,n}$ and $V_{m,n}$ can be constructed by the method in Ref.~\cite{eddins2021doubling}.
Those circuits are required to produce the superposition of computational basis states $\ket{m}$ and $\ket{n}$ (note that $U_{m,n}^\dag\ket{0} = (\ket{m}+\ket{n})/\sqrt{2}$ and $V_{m,n}^\dag\ket{0} = (\ket{m}-i\ket{n})/\sqrt{2}$ ).
Such a superposition can be realized by combining $\mr{Hamming}(m,n)$ CNOT gates and several single-qubit gates, where $\mr{Hamming}(m,n)$ is the Hamming distance between the binary representations of integers $m$ and $n$.
In short, at most $N$ CNOT gates are needed to construct the circuits $U_{m,n}$ and $V_{m,n}$.

\subsubsection{Explicit algorithm}

Having all the ingredients for the expectation value $\ev{O}{\psi}$ in Eq.~\eqref{eq:ev_O_3} explained, we are now in the position to describe the explicit procedure of our algorithm.

Before going into the actual procedure, we introduce some premise of the algorithm.
Let $R$ be a number of the computational basis states kept in the sum.
Concretely, we first approximate the original state $\ket{\psi}$ of Eq.~\eqref{eq:state_1} by
\begin{align}
\ket{\psi} \approx \ket{\psi_R}
= \mc{N}_R \sum_{r=1}^{R}
   \bra{z_r}\ket{\psi} \ket{z_r},
\label{eq:state_2}
\end{align}
where $z_1, \cdots, z_R \in \{0,1,\cdots,2^N-1\}$
represent labels of the computational basis states that have $R$ most-significant weights in descending order, $|\ip{z_1}{\psi}|^2 \geq \cdots \geq |\ip{z_R}{\psi}|^2$, and 
$\mc{N}_R = 1 / \sqrt{\sum_{r=1}^R \abs{\bra{z_r}\ket{\psi}}^2}$ is the normalization factor.
Eq.~\eqref{eq:ev_O_3} under this approximation becomes
\begin{align}
 \ev{O}{\psi} &\approx \ev{O}{\psi_R} \notag \\
 &= {\mc{N}_R}^2 \sum_{r,\rp=1}^{R}
    \abs{\bra{z_r}\ket{\psi}}^2\abs{\bra{z_{\rp}}\ket{\psi}}^2
    \frac{\mel{z_r}{O}{z_{\rp}} }{\bra{z_r}\ket{\psi} \bra{\psi}\ket{z_{\rp}}}.
\label{eq: ev_approx}
\end{align}

Picking up $z_1,..,z_R$ and evaluating their single weight factors $|\ip{z_r}{\psi}|^2$ as well as the normalization factor $\mc{N}_R$ can be done solely based on a single type of projective measurements of $\ket{\psi}$ in the computational basis, as explicitly prescribed later.
The transition matrix elements $\mel{z_r}{O}{z_{\rp}}$ can be efficiently calculated by classical computers.
The interference factors $\ip{z_r}{\psi} \ip{\psi}{z_{\rp}}$, for off-diagonal $r\neq \rp$ elements, can be evaluated based on Eq.~\eqref{eq:intf_factor_1}, which requires the runs of $\mc{O}(R^2)$ distinct quantum circuits in addition to those for single weight factors, naively, as explained in the precious subsection.
However, by using an equality
\begin{align}
\bra{z_r}\ket{\psi}\bra{\psi}\ket{z_{\rp}}
= \frac{(\bra{\ell}\ket{\psi}\bra{\psi}\ket{z_r})^*\bra{\ell}\ket{\psi}\bra{\psi}\ket{z_{\rp}}}{\abs{\bra{\ell}\ket{\psi}}^2}
\label{eq:intf_factor_2}
\end{align}
for $\ell \neq z_r, z_{\rp}$ satisfying $\bra{\ell}\ket{\psi}\neq 0$, one can reduce the number of the required circuits to $2(R-1) = \mc{O}(R)$.
That is, one can reconstruct all the off-diagonal elements of $\ip{z_r}{\psi} \ip{\psi}{z_{\rp}}$ based on a subset of interference factors $\{ \bra{\ell}\ket{\psi}\bra{\psi}\ket{z_r} \}_{z_r\neq \ell}$ and the single-weight factor $\abs{\bra{\ell}\ket{\psi}}^2$ for some fixed $\ell \in \{z_1,\cdots,z_R \}$.
For example, choosing $\ell = z_1$, one can calculate all the interference factors $\ip{z_r}{\psi} \ip{\psi}{z_{\rp}}$ from 
the set $\{ \bra{z_1}\ket{\psi}\bra{\psi}\ket{z_2}, \cdots, \bra{z_1}\ket{\psi}\bra{\psi}\ket{z_R} \}$
that requires $2(R-1)$ distinct quantum circuits to evaluate, besides one quantum circuit for the single-weight factors $\abs{\bra{z_r}\ket{\psi}}^2$.
(See Sec.~\ref{subsec:improve} and Appendix~\ref{appsec: interference} for possible improvements.)
After all the elementary quantities (single weight factors, normalization factor, transition matrix elements, and interference factors) are evaluated, summing up all the terms in Eq.~\eqref{eq: ev_approx} by classical computers yields the approximate estimate of the expectation value $\ev{O}{\psi}$.

The following is the actual procedure of our algorithm.

\noindent {\bf Algorithm:}
\begin{enumerate}
\item Prepare the state $\ket{\psi}$ on quantum computers, followed by the measurement in the computational basis $\ket{n}$ ($n=0,1,\cdots,2^N -1$). Repeat the measurement $L_f$ times resulting in a sequence of outcomes $\qty{x} = x^{(1)},x^{(2)}, \cdots, x^{(L_f)}$, where $x^{(i)} \in \{ 0,1,\cdots,2^N -1\}$.
\item Pick up the most frequent $R$ elements from $\qty{x}$, and sort them into descending order of frequency, leading to a rearranged sequence of $\qty{z} = z_1, z_2, \cdots, z_R$, where $z_r \in \{0,1,\cdots,2^N -1\}$ and $z_r \neq z_{\rp}$ for $r \neq \rp$.
Suppose $z_r$ appears $T_r$ times in $\qty{x}$.
Then, quantities defined by
\begin{align}
f_r &\equiv \frac{T_r}{L_f} \quad
(r=1,2,\cdots, R), \\
c_R &\equiv \frac{1}{\sqrt{ \sum_{r=1}^R T_r / L_f }} ,
\end{align}
provide estimates for the single-weight factors and the normalization factor, i.e., $f_r \simeq \abs{\bra{z_r}\ket{\psi}}^2$ and $c_R \simeq \mc{N}_R$, respectively.
\item Evaluate relevant transition matrix elements $\mel{z_r}{O}{z_{r'}}$ ($r,r' = 1,2,\cdots, R$) by classical computations.
\item For $r = 2,\cdots, R$, prepare a quantum state $U_{z_1, z_{r}}\ket{\psi}$, followed by the measurement in the computational basis;
repeat the measurement $L_{A_r}$ times and count the occurrence of the outcome ``0".
When ``0" appears $T^{(A)}_r$ times, construct
\begin{align}
A_r &\equiv \frac{T^{(A)}_r}{L_{A_r}}\\
&\simeq \abs{\mel{0}{U_{z_1,z_r}}{\psi} }^2 = \abs{\frac{\bra{z_1} + \bra{z_r}}{\sqrt{2}} \ket{\psi}}^2,  \notag
\end{align}
which is equivalent to $\mathcal A_{z_1, z_r}$.
In a similar way, estimate the quantity equivalent to $\mathcal B_{z_1, z_r}$:
\begin{align}
B_r &\equiv \frac{T_r^{(B)}}{L_{B_r}}\\
&\simeq \abs{\mel{0}{V_{z_1,z_r}}{\psi} }^2 = \abs{\frac{\bra{z_1} + i\bra{z_r}}{\sqrt{2}} \ket{\psi}}^2 \notag
\end{align}
by repeating $L_{B_r}$ measurements for the state $V_{z_1, z_r}\ket{0}$ in the computational basis and obtaining the outcome ``0" $L^{(B)}_r$ times.
\item Combine $f_r, A_r, B_r$ on classical computers to estimate interference factors.
Equation~\eqref{eq:intf_factor_1} now reads
\begin{align}
g_r &\equiv A_r +iB_r -\frac{1+i}{2}(f_1 +f_r) \label{eq:intf_factor_3}\\
&\simeq \bra{z_1}\ket{\psi}\bra{\psi}\ket{z_r} \notag
\end{align}
for $r=2,\cdots, R$.
The rest of the off-diagonal elements, i.e., $\bra{z_r}\ket{\psi}\bra{\psi}\ket{z_\rp}$ for $r,\rp = 2,\cdots,R$ with $r\neq \rp$, are estimated by Eq.~\eqref{eq:intf_factor_2}, now recast as
\begin{align}
G_{r,r'} \equiv \frac{g_r^* g_{r'} }{f_1} \simeq \bra{z_r}\ket{\psi}\bra{\psi}\ket{z_{r'}}.
\end{align}
The diagonal elements of the interference factors are $\abs{\bra{z_r}\ket{\psi}}^2$, which are already obtained.
\item Estimate the expectation value by substituting the quantities in Eq.~\eqref{eq: ev_approx} with $f_r, c_R, A_r, B_r, g_r$ and $G_{r,r'}$, accordingly.
\end{enumerate}

Before ending this section, let us discuss the number of distinct quantum circuits to be measured in our algorithm.
We need one quantum circuit to obtain $\{ f_r \}$ and $2(R-1)$ quantum circuits to obtain $A_r, B_r (r=2,\cdots,R)$;
hence the total number of the distinct quantum circuits is $2R-1$~\footnote{
We note that this is consistent with the degrees of freedom in the truncated state $\ket{\psi_R}$ introduced by Eq.~\eqref{eq:state_2}.
We assume the state is pure and hence the degrees of freedom counted in terms of real coefficients are $2R-1$ (``$-1$'' comes from the normalization).
}.
In contrast, the conventional method to estimate expectation values, in the simplest implementation, requires $M$ quantities (i.e., distinct quantum circuits) to be measured with $M$ being the number of Pauli strings in the expansion of the observable $O$.
We therefore expect that our algorithm is more efficient when $R$ is significantly smaller than $M$, i.e., when the state is well-concentrated. 
Yet, this comparison is obviously not enough to show efficiency of our algorithm.
For instance, in the conventional method, Pauli strings with small coefficients could be truncated in the expansion of the observable in  Eq.~\eqref{eq:O_pauli} on similar footing as truncating the expansion of the wavefunction in Eq.~\eqref{eq:state_2}, which would reduce the number of quantities to be measured.
We can effectively perform such a truncation by optimizing the measurement allocation. We can also do so for our proposed algorithm to effectively reduce the number of quantities to be measured.
Besides, in the conventional method, grouping simultaneously-measurable Pauli strings can further reduce the number of quantities to be measured.
Therefore, in the next section, we numerically evaluate statistical fluctuations of the expectation values estimated in the conventional and our methods, to discuss efficiency.

\section{Numerical comparisons of statistical fluctuations for small molecules \label{sec:experiment}}

In this section, we evaluate performance of our method in comparison with the conventional methods to estimate the expectation values by taking Hamiltonians of small molecules as examples.
First, we numerically examine whether and how ground states of those Hamiltonians are concentrated in the computational basis, where each of the basis states corresponds to a Slater determinant state.
Second, we evaluate variances of the estimated expectation values originating from finite numbers of measurements.
Then, we estimate the required number of measurements to ensure the standard deviation is as small as $10^{-3}$ Hartree, with the chemical accuracy in mind, for each molecule.

\subsection{Setup for numerical analysis
}

\begin{table*}
\caption{
Molecules adopted in our numerical study.
The number of qubits and the number of Pauli strings ($M$) contained in the Hamiltonian are shown for each molecule.
$R$ is the number of computational basis states retained in the approximate expectation value (Eq.~\eqref{eq: ev_approx}) for the exact ground state.
The number of quantities to be measured is $M$ in the simplest implementation of the conventional method, while is $\mathcal{O}(R)$ in our algorithm.
The reduction of this number suggests an efficacy of our method, although a quantitative comparison with numerical evaluation of variances is necessary as performed in Sec.~\ref{sec:experiment}.
$E_R- E_{\rm exact}$ is the difference between the energy expectation values for the truncated state and the exact ground state. 
}
\begin{center}
\begin{tabular}{l|cccccccc}
\hline\hline
Molecule & Qubits & Pauli terms & $R$ & $E_R - E_{\rm exact}$ [$10^{-4}$ Hartree] \\
\hline
\ce{H_2} & $4$ & $15$ & $2$ & $0$ \\
\ce{LiH} & $12$ & $631$ & $9$ & $2.4$ \\
\ce{H2O} & $14$ & $1086$ & $30$ & $2.8$ \\
\ce{NH3} & $16$ & $3057$ & $171$ & $4.4$ \\
\ce{CH4} & $18$ & $2212$ & $322$ & $4.2$ \\
\ce{CO} & $20$ & $4427$ & $379$ & $5.4$ \\
\ce{H2S} & $22$ & $6246$ & $31$ & $6.4$ \\
\ce{C2H2} & $24$ & $5185$ & $1556$ & $5.0$ \\
\hline\hline
\end{tabular}
\end{center}
\label{tab:info}
\end{table*}

To see the performance of our method, we consider various molecular Hamiltonians for electronic states under the Born-Oppenheimer approximation and their ground states with 4 to 24 spin orbitals, listed in Table~\ref{tab:info}.
The second-quantized electronic Hamiltonians of those molecules are generated by OpenFermion~\cite{McClean_2020} interfaced with PySCF~\cite{sun2018pyscf,sun2020recent}.
We employ the Hartree-Fock orbitals with the STO-3G minimal basis set as spin orbitals and use the Jordan-Wigner transformation to map fermionic Hamiltonians into qubit ones (see a review~\cite{mcardle2018quantum} for technical details). 
We take a point group symmetry of a molecule into account in the Hartree-Fock calculation when the symmetry is present.
In this setup, each of the computational basis states corresponds to a single Slater determinant state.
The molecular geometries are chosen to match Refs.~\cite{hadfield2020measurements, wu2021overlapped} except for \ce{CH4}, CO, \ce{H2S}, and \ce{C2H2}.
The geometries for these four molecules are taken from CCCBDB database by NIST~\cite{cccbdb}.
The detailed information on the geometries is given in Table~\ref{tab: geometries} in Appendix~\ref{appsec: numerics}.
We calculate the exact ground states of the Hamiltonians by the full-configuration interaction (FCI, or exact diagonalization) method and take it as the state $\ket{\psi}$ to evaluate the variances of the expectation values of the Hamiltonians.
A numerical library Qulacs~\cite{suzuki2021qulacs}, a fast simulator of quantum circuits, is used for calculations of expectation values like $\ev{P_i}{\psi}$.

The amplitudes $\bra{n}\ket{\psi}$ are real for all the molecular ground states $\ket{\psi}$ we consider.
As such, the estimated energy expectation values are insensitive to $B_r$, which enter as imaginary part of $\bra{z_1}\ket{\psi}\bra{\psi}\ket{z_r}$ (see Eq.~\eqref{eq:intf_factor_3}), and hence one does not need to perform measurements to estimate $B_r$.
In this case, $R$ distinct quantum circuits are required in total for our algorithm.
In the expressions below, we still keep possible contributions from $B_r$ for the purpose of generality.

\subsection{
Concentration in computational basis and error caused by truncation
\label{subsec: wfn analysis}}

\begin{figure*}
\includegraphics[width=0.45\textwidth]{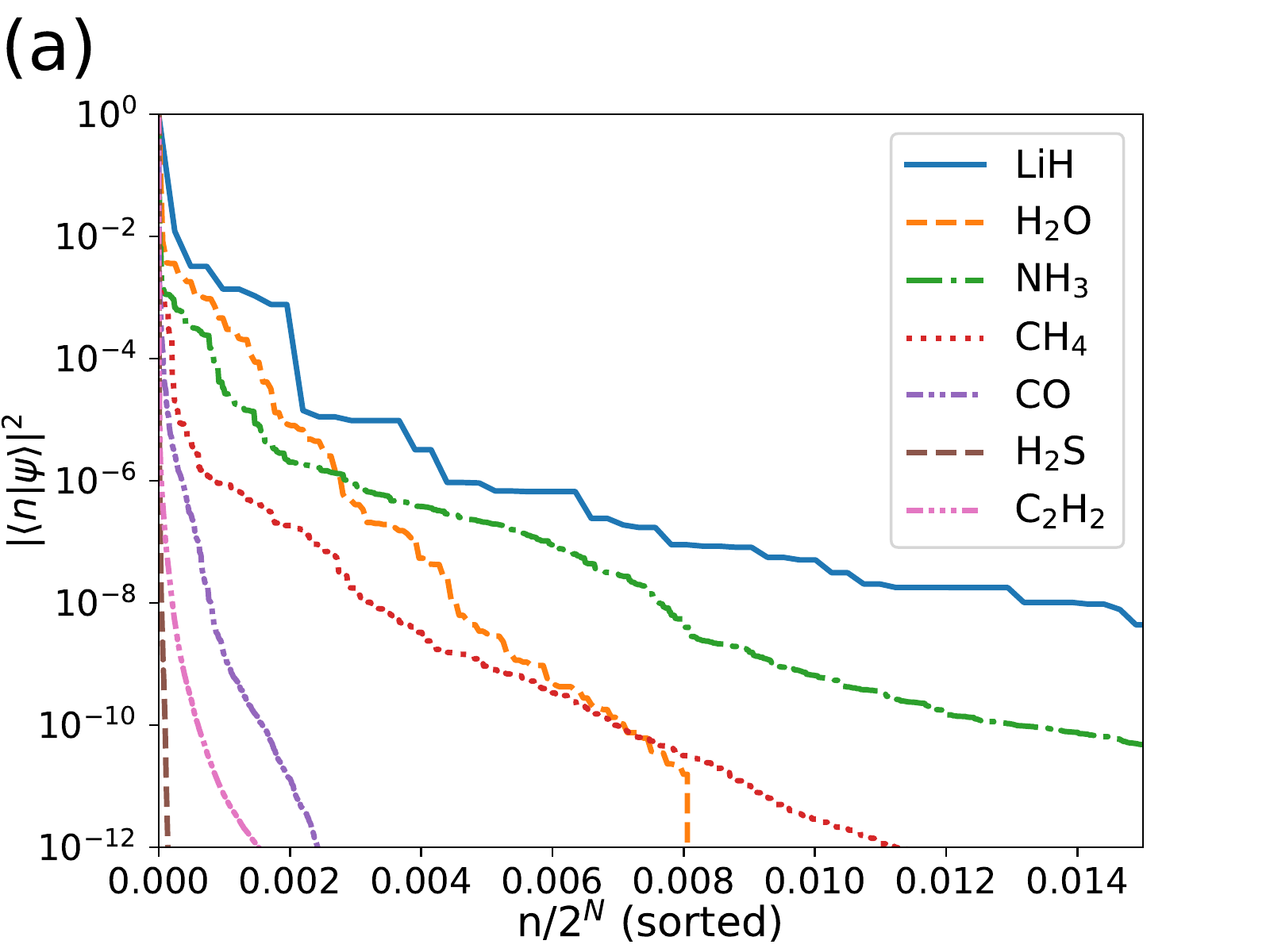}
\includegraphics[width=0.45\textwidth]{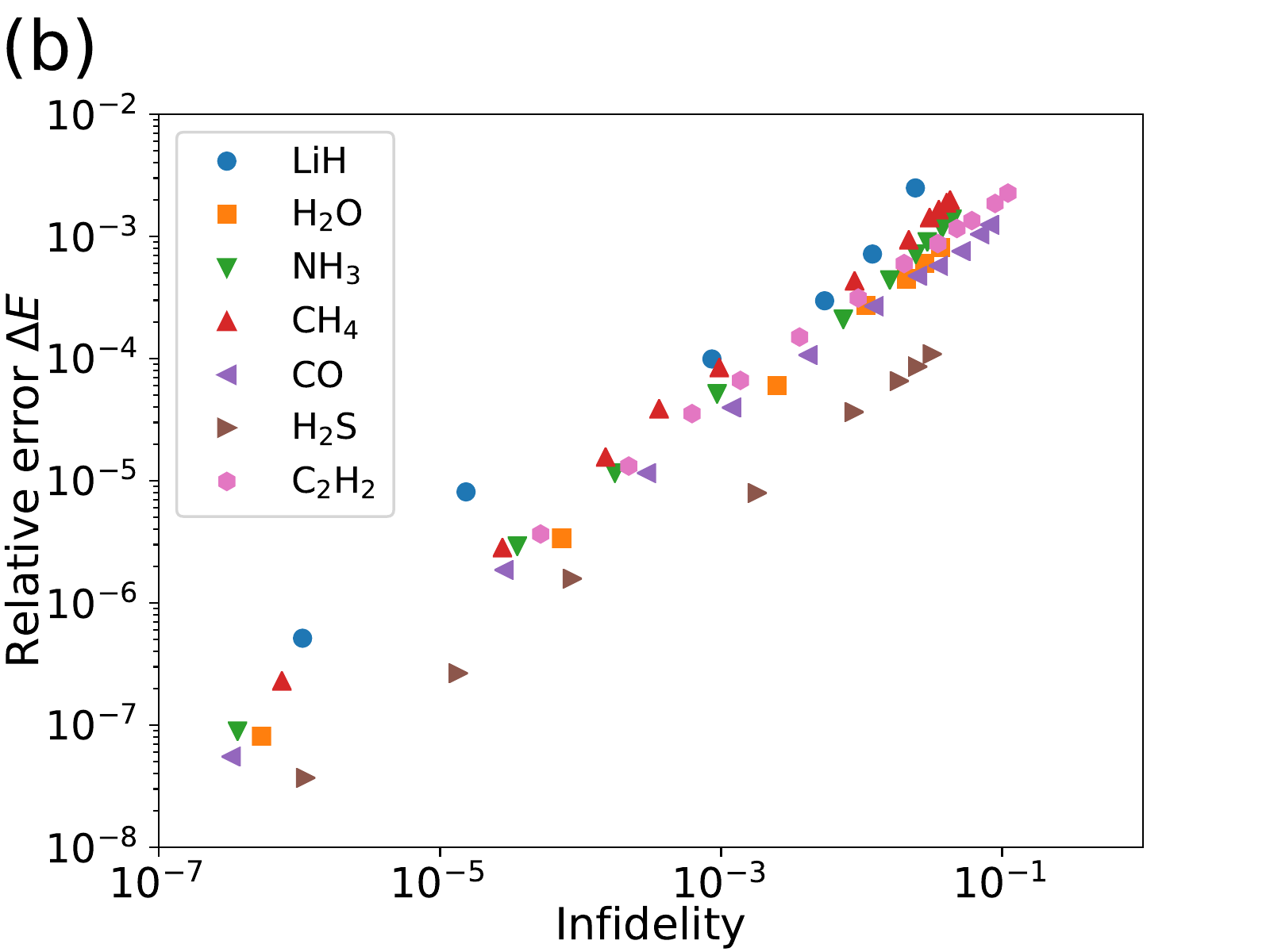}
\caption{
(a) Concentration in the computational basis shown for ground states $\ket{\psi}$ of molecular Hamiltonians listed in Table~\ref{tab:info}.
The distribution of $\abs{\bra{n}\ket{\psi}}^2$ is plotted for $n/2^N$, where $\ket{n}$ is a computational basis state and $N$ is the number of qubits.
The index $n$ of the computational basis is sorted in descending order of $\abs{\bra{n}\ket{\psi}}^2$.
(b) The relative error vs infidelity plot for points corresponding to various $R$ values of different molecular ground states.
The relative error $\Delta E = (E_R - E_\mr{exact})/|E_\mr{exact}|$ is defined for the energy expectations values of the exact ground state $\ket{\psi}$ and truncated state $\ket{\psi_R}$ of Eq.~\eqref{eq:state_2}.
The infidelity is given by $1 -|\ip{\psi_R}{\psi}|^2$.
See the text for detailed explanation.
\label{fig:dist and error}
}
\end{figure*}

We first see whether and how actual ground states of the molecular Hamiltonians are concentrated in the computational basis. Fig.~\ref{fig:dist and error}(a) shows the distribution of the single-weight factors $\abs{\bra{n}\ket{\psi}}^2$ in the computational basis ($n = 0,1,\cdots, 2^N -1$) sorted in descending order, for each of molecular ground states.
The basis index $n$ is normalized by the total number of basis states $2^N$ for comparison between molecules of different numbers of qubits $N$.
We observe the distributions decay quickly with $n/2^N$, implying the ground states are concentrated.
Hence we expect only a small fraction of the basis states are relevant in the estimation of the energy expectation values for those molecules.

Next, we investigate the error caused by truncating less-significant computational basis states in the calculation of the expectation value.
As explained in the previous section, we retain $R$ most-significant computational basis states by truncating the double sum in the expectation value formula of Eq.~\eqref{eq:ev_O_3}, and adopt Eq.~\eqref{eq: ev_approx} as an estimator for the expectation value.
One must choose a value of $R$ so that the error caused by this truncation is within a required accuracy of the expectation value.
Here we present a criterion for this, guided by a formal argument in Appendix~\ref{app:fidelity}.

As shown in Eq.~\eqref{appeq:bound}, the difference between expectation values of an observable $O$ for the original state $\ket{\psi}$ and approximate state $\ket{\psi_R}$ is bounded by their fidelity:
\begin{align}
 |\ev{O}{\psi} - \ev{O}{\psi_R} | \leq 2 \|O\|_{\infty} \sqrt{1 - |\ip{\psi_R}{\psi}|^2 },
\label{eq:bound}
\end{align}
where $\|O\|_\infty$ is the largest singular value of $O$ and $|\ip{\psi_R}{\psi}|^2$ is the fidelity between the two states. 
The fidelity can be expressed as
\begin{align}
\abs{ \bra{\psi_R}\ket{\psi} }^2 = \sum_{r=1}^R \abs{\bra{z_r}\ket{\psi}}^2,
\end{align}
and hence can be estimated by $1/c_R^2 = \sum_{r=1}^R T_r / L_f$ in our algorithm.
This does not require additional measurements.
Therefore we are led to take the estimated fidelity $|\ip{\psi}{\psi_R}|^2$ as a quick indicator to determine $R$.

In the application to the molecular Hamiltonian, our interest is constraining the difference of energy expectation values $E_\mr{exact} = \ev{H}{\psi}$ and $E_R = \ev{H}{\psi_R}$, for the exact ground state $\ket{\psi}$ and truncated state $\ket{\psi_R}$.
In Fig.~\ref{fig:dist and error}(b), the relative error $\Delta E = (E_R - E_\mr{exact})/|E_\mr{exact}|$ and the infidelity $1 - |\ip{\psi_R}{\psi}|^2$ are plotted for various $R$ values of different molecules.
The relative error and infidelity seem positively correlated, and we observe $\Delta E$ is a factor of $0.1$--$0.01$ smaller than the infidelity depending on molecular species.
It is notable that this dependence is rather mild.
Based on this result, we adopt a constraint on the infidelity as the concrete criterion for determining $R$ in our numerical analysis, which is given as follows: we choose the minimum integer $R'$ which satisfies $1 - |\ip{\psi_{R'}}{\psi}|^2 \leq 10^{-4}$ as $R$, for all the molecules in our study.
In Table~\ref{tab:info}, the values of $R$ determined by this procedure are shown with the (absolute) truncation error $E_R - E_{\rm exact}$.
All the molecules exhibit the error smaller than $10^{-3}$ Hartree.

\subsection{Comparisons of statistical fluctuations with conventional methods
\label{subsec: shot experiment}}
\begin{figure*}
\includegraphics[width=0.49\textwidth]{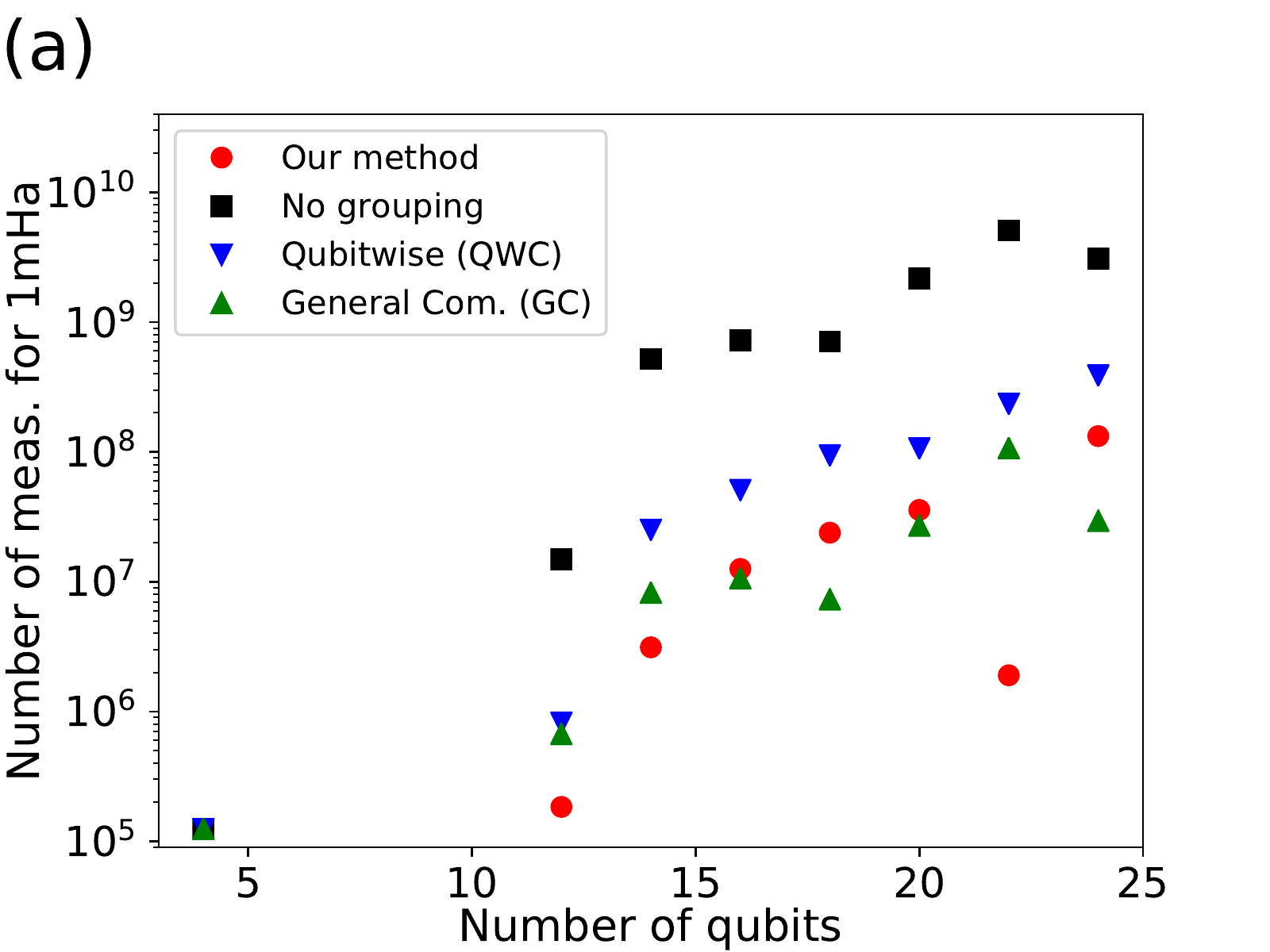} 
\includegraphics[width=0.49\textwidth]{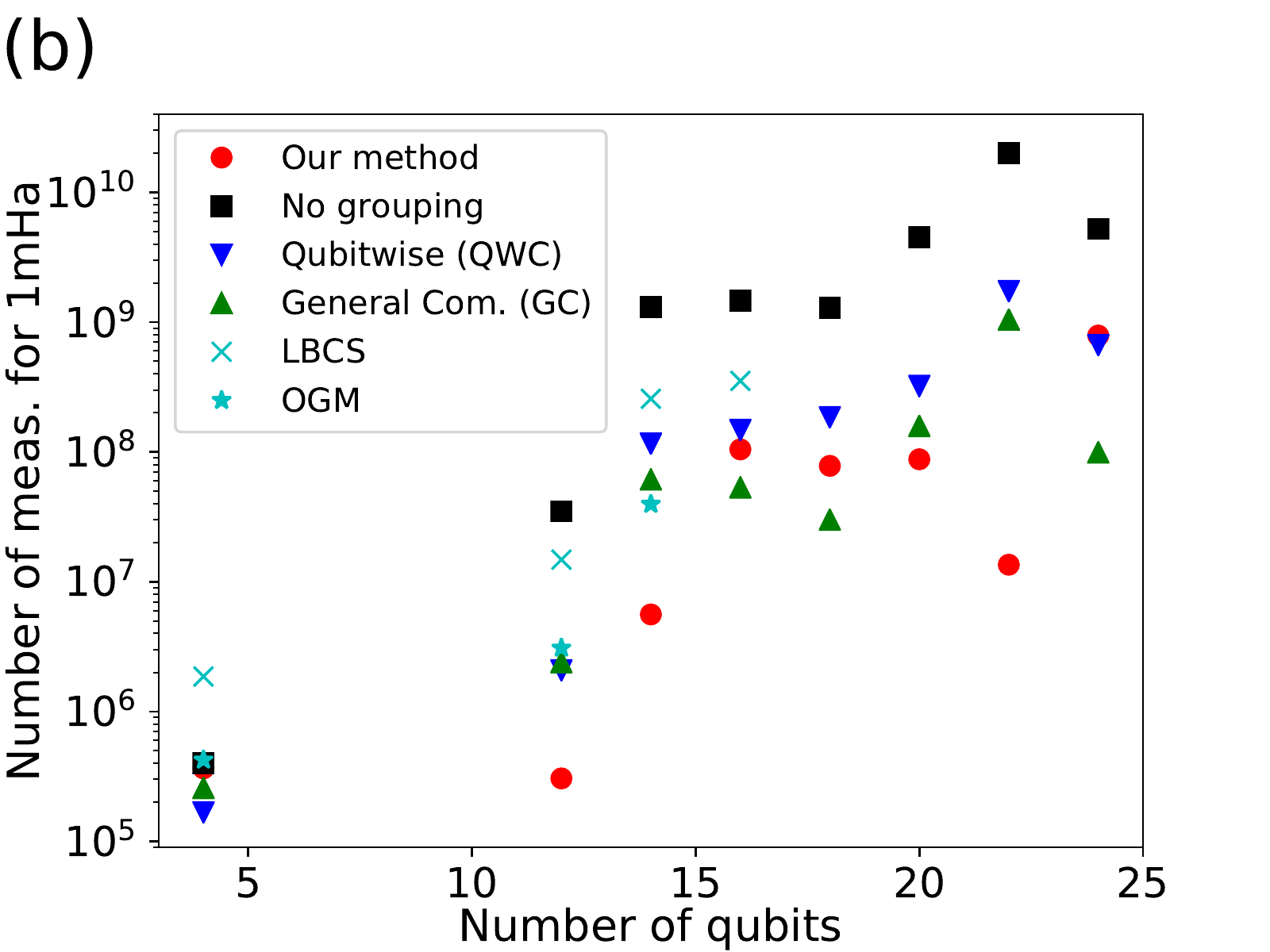} 
\caption{
The total number of measurements required to achieve the standard deviation of $10^{-3}$ Hartree in the energy expectation value estimation with the ground state for each of molecules listed in Table~\ref{tab:info}, based on a different estimation method and measurement allocation strategy.
(a) The results based on the measurement allocations optimized for the exact ground states (obtained by FCI calculations).
(b) The results based on the heuristic measurement allocations.
LBCS and OGM stand for locally-biased classical shadow~\cite{hadfield2020measurements} and overlapped grouping measurement~\cite{wu2021overlapped}, respectively.
See the text for detailed explanation.
\label{fig:shot}
}
\end{figure*}

We estimate statistical fluctuations of expectation values estimated by our method and compare them with those by the other existing methods.

As for our method, we formulate the variances of estimated expectation values in Appendix~\ref{appsubsec: variance ours},
where the details can be found.
We here consider only the statistical fluctuations caused by finite numbers of measurements $L_f, \{L_{A_r}\}$, and $\{L_{B_r}\}$ for $f_r, A_r$ and $B_r$, respectively, and do not include any other sources of fluctuations and deviations such as coherent noises in quantum devices.
Then, the variance of an expectation value estimated by our method, Eq.~\eqref{eq: ev_approx}, can be expressed in terms of the numbers of measurements as
\begin{align}
\mr{Var}[\ev{O}{\psi}]
= \frac{v_f}{L_f}
+\sum_{r=2}^R \pqty{
 \frac{v_{A_r}}{L_{A_r}} +\frac{v_{B_r}}{L_{B_r}}},
\label{eq:var_O}
\end{align}
where the coefficients $v_f$, $v_{A_r}$ and $v_{B_r}$ are defined by Eq.~\eqref{eq: variance of our method_where}.
The total number of measurements is
\begin{align}
L = L_f +\sum_{r=2}^R\pqty{L_{A_r} +L_{B_r}}.
\end{align}

We also investigate how to optimally allocate the measurement budget to each of $L_f, \{L_{A_r}\}$, and $\{L_{B_r}\}$, 
when the total number of measurements $L$ is fixed to a certain value.
That is, we consider the minimization of the variance~\eqref{eq:var_O}, under the condition of a fixed $L$.
We consider two cases:
the first case is when one knows true values of coefficients $v_f$, $v_{A_r}$ and $v_{B_r}$ appearing in the variance formula~\eqref{eq:var_O};
the second case is when one cannot access those values.
In the first case, the optimal allocation of measurements can be achieved and is given by
\begin{align}
L_f:L_{A_r}:L_{B_r}=\sqrt{v_f}:\sqrt{v_{A_r}}:\sqrt{v_{B_r}},
\label{eq:shot-allocation}
\end{align}
recasting the variance formula into Eq.~\eqref{eq: best variance ours}.
In a practical situation where the algorithm is applied to a general state, however, it would be unrealistic to assume values of $v_f$, $v_{A_r}$ and $v_{B_r}$ are known before the measurements are conducted, as those values depend on quantities such as $f_r$, $A_r$ and $B_r$ (see Eq.~\eqref{eq: variance of our method_where}), which are supposed to be constructed based on the measurement outcomes.
Hence, as a more practical case, we also consider the second case.

In the second case, where we are ignorant of values of $v_f$, $v_{A_r}$ and $v_{B_r}$ for a given state, we adopt a heuristic method: we take a reference state for which values of $v_f$, $v_{A_r}$ and $v_{B_r}$ are classically calculated prior to the measurements, and optimize the measurement allocation for this state by Eq.~\eqref{eq:shot-allocation};
then, we apply the same allocation to arbitrary states considered in the algorithm.
In this article, we take the following state as the reference state:
\begin{align}
  \ket{\psi_w} = \sqrt{w} \ket{z_1} + \sum_{r=2}^R \sqrt{\frac{1-w}{R-1}}\ket{z_r},
\label{eq:heuristic}
\end{align}
where $w$ is a positive real parameter with $0<w<1$.
This is designed as a simplified representative for concentrated states.
Using the measurement allocation optimized for $\ket{\psi_w}$ with some $w$, we estimate the variance for a given state in the second case (see Eqs.~\eqref{eq: shots for heuristic variance ours} and \eqref{eq: heuristic variance ours} for explicit formulas).
The formulas with detailed derivations are presented in Appendix~\ref{appsubsec: variance ours}.

As for the conventional methods, we consider the methods based on the decomposition of an observable into Pauli strings with or without partitioning into co-measurable groups, as briefly described in Sec.~\ref{subsec: conventional methods}. 
The way of grouping of Pauli strings and the choice of the measurement allocation to each group both affect
the variance of the estimated expectation value.
We examine combinations of three types of Pauli-string grouping and two types of measurement allocation to each of grouped Pauli strings.
For the grouping, we consider (i) no grouping, i.e., all the Pauli strings in an observable are separately measured, (ii) qubit-wise commuting (QWC) grouping, and (iii) general-commuting (GC) grouping.
The QWC grouping requires no additional two-qubit gates to simultaneously measure all the Pauli strings in each group~\cite{kandala2017, verteletskyi2020}, while the GC grouping needs extra $\mc{O}(N^2/\log N)$ two-qubit gates~\cite{yen2020, aaronson2004improved} (see also \cite{gokhale2020ON3,crawford2021efficient}).
The details of the three grouping methods are summarized
in Appendix~\ref{appsec: grouping}.
For the measurement allocation, we again consider the situation where one knows true expectation values of Pauli strings such as $\ev{P_i}{\psi}$ prior to the measurements, and the situation where one cannot access those values without the measurements, just as in our method.
The measurement allocation is optimized for the state $\ket{\psi}$ under consideration in the former, while is optimized for Haar random states in the latter.
The variances for those two ways of allocation are given by Eqs.~\eqref{eq: best variance conventional fci} and \eqref{eq: best variance conventional haar}, respectively (see Appendix~\ref{appsubsec: variance conventional} for details).

Besides the conventional methods, we also consider two recently-proposed methods based on a technique called classical shadow for comparison:
one is locally-biased classical shadow (LBCS)~\cite{hadfield2020measurements} and the other is overlapped grouping measurement (OGM)~\cite{wu2021overlapped}.
These methods work without knowing expectation values of observables before measurements and hence are compared with our heuristic method of the measurement allocation.
We take the values of the variances presented in Refs.~\cite{hadfield2020measurements, wu2021overlapped} for \ce{H2}, \ce{LiH}, \ce{H2O} (LBCS and OGM) and \ce{NH3} (only LBCS).

In Fig.~\ref{fig:shot}, we show the total number of measurements $L$ required to achieve the standard deviation of $10^{-3}$ Hartree, in energy expectation value estimation with the ground state (obtained by the FCI calculation) for each of molecules listed in Table~\ref{tab:info}, based on a different estimation method and measurement allocation strategy explained above.
In Fig.~\ref{fig:shot}(a) the measurement allocations are optimized for the exact ground states in both of the conventional and our methods, while in Fig.~\ref{fig:shot}(b) they are optimized for some reference states, i.e., for the Haar random states in the conventional methods and for the heuristic reference states (Eq.~\eqref{eq:heuristic}) in our method.
As the variances can be parameterized as $c_v/L$ for a total number of measurements $L$ with some measurement allocation (see Eq.~\eqref{eq: general optimal variance general}), one may infer the required $L$ for $10^{-3}$ Hartree by
\begin{align}
L = \qty(\frac{\sqrt{c_v}}{10^{-3}~\ce{Hartree}})^2,
\label{eq:shot_1mHa}
\end{align}
where $c_v$ is a constant defined by Eq.~\eqref{eq: general optimal variance general} with Eqs.~\eqref{eq: general optimal variance} and \eqref{eq: general optimal variance ref}.
We take $w=0.75$ in Eq.~\eqref{eq:heuristic} for our heuristic way of the measurement allocation.
In both cases of the measurement allocation strategies, we observe that our method tends to require a smaller number of measurements than two conventional methods, no grouping and QWC grouping.
In contrast, the GC grouping tends to compete with our method.
Nevertheless, this may still indicate an advantage of our method, given gate complexity of two methods: that is, our method needs $\mc{O}(N)$ two-qubit gates while the GC grouping requires $\mc{O}(N^2/\log N)$ two-qubit gates to estimate an expectation value, in addition to gates for constructing a circuit of the state $\ket{\psi}$ of interest.
We note that both of our method and GC grouping tend to demand fewer measurements than the classical-shadow based approaches, LBCS and OGM.

The results in Fig.~\ref{fig:shot} support the effectiveness of our method in estimating expectation values of molecular Hamiltonians for their ground states.

\section{Discussion
\label{sec:discussion}}

We comment on several points in our study before summarizing this article.

\subsection{
Potential advantages of our method over conventional methods}

Our method performs the summation in Eq.~\eqref{eq: ev_approx}, which has $R^2$ terms, by classical computers.
In general, a value of $R$ that is large enough to approximate the expectation value with sufficient accuracy may be exponentially large in the number of qubits $N$, which will deteriorate the efficiency of our method.
The numerical analysis in the previous section demonstrates
examples in molecular systems where $R$ can be taken small enough to efficiently process the summation in Eq.~\eqref{eq: ev_approx}
with sufficient accuracy retained.
Although the scaling of $R$ with $N$ for larger molecules is not known, we expect that values of $R$ do not become so large for ground states of weakly-correlated molecules.
In Appendix~\ref{appsec: N2 result}, we also investigate the effect of electron correlation by performing the same numerical analysis to the nitrogen molecule \ce{N2} for various bond lengths.
As \ce{N2} is stretched from the equilibrium bond length and the triple bond is broken, 
the electron correlation is considered to be large, but our method is still more effective than the QWC grouping method.
Our numerical results give concrete examples in which our method realizes an actual speedup over the conventional methods in expectation value estimation of observables.

Let us discuss the noise robustness of our method.
Since our method picks up dominant computational basis states in the first step of the algorithm, it is easy to mitigate errors of quantum devices that change a symmetry of the computational basis states, by the method called symmetry verification~\cite{bonet2018,mcardle2019error,huggins2021efficient}.
For example, when we treat an electronic Hamiltonian of some molecule conserving the particle number and employ its qubit representation in which each computational basis state has a definite particle number, e.g., the Jordan-Wigner transformation, the bit-flipping error that changes the count of ``1" in the binary representation of the basis state can be easily detected; the ground state must have some fixed particle number, i.e., the number of electrons, and we can simply reject (if any) an observed basis state $\ket{z_r}$ that does not have such a particle number. 
Other types of noises such as dephasing error can be mitigated by the standard techniques of quantum error mitigation~\cite{temme2017,endo2018} in the same manner as the conventional methods.

\subsection{Improvements and extensions of our method
\label{subsec:improve}}

We discuss possible improvements and extensions of our algorithm.
First, it may be possible to calculate the interference factors $\bra{m}\ket{\psi}\bra{\psi}\ket{n}$ or $\bra{z_r}\ket{\psi}\bra{\psi}\ket{z_{r'}}$ in different ways than our explicit algorithm described in Sec.~\ref{subsec:our-method}, 
with lower quantum computational costs.
We discuss possible directions in Appendix~\ref{appsec: interference}.
For instance, if $\mr{Hamming}(m,n) = 1$, where $\mr{Hamming}(m,n)$ is the Hamming distance between the binary representations of integers $m$ and $n$, all the interference factors $\bra{m}\ket{\psi}\bra{\psi}\ket{n}$ for possible combinations of arbitrary $m$ and $n$ can be evaluated just based on $2N$ distinct quantum circuits, given the single-weight factors $\abs{\bra{m}\ket{\psi}}^2$ and $\abs{\bra{n}\ket{\psi}}^2$ are already obtained.
Each of the quantum circuits only requires an application of the Hadamard (and phase) gates on the state $\ket{\psi}$.
Similarly, if $\mr{Hamming}(m,n) = 2$, all the possible interference factors $\bra{m}\ket{\psi}\bra{\psi}\ket{n}$ can be evaluated just based on $N(N-1)$ distinct quantum circuits, each of which requires a consecutive application of the CNOT and Hadamard (and phase) gates on the state $\ket{\psi}$.
These procedures combined with the relation~\eqref{eq:intf_factor_2} may lead to an improved way to evaluate the interference factors which requires a smaller number of distinct quantum circuits.
See Appendix~\ref{appsec: interference} for details.

Second, in the application to quantum chemistry, fermionic states other than Slater determinants may be identified as the computational basis states, so long as they meet the requirements such as concentration of the state and feasibility of efficient calculation for the transition matrix elements in the basis states.
For example, the configuration state functions (CSFs)~\cite{helgaker2014molecular,jensen2017introduction}, which preserve the particle number, the total spin, etc., may be identified as the computational basis states, given transition matrix elements between two CSFs can be classically evaluated in $\mc{O}(\mr{poly}(N))$ time.
In the conventional methods to evaluate expectation values, the number of Pauli strings in a Hamiltonian written in such a basis is larger than that written in a basis of Slater determinants and it becomes more costly to evaluate the expectation value.
On the other hand, in our method, putting more focus on a quantum state, the expectation value will be evaluated efficiently as long as the state is concentrated.

\subsection{Relationship to classical computational methods}

Finally, we would like to discuss relationship between our algorithm and classical computational methods in quantum chemistry.
Performing the importance sampling for Eq.~\eqref{eq:ev_O_3} is considered in the variational Monte Carlo (VMC)~\cite{sabzevari2018faster,mahajan2020efficient}, although the summations for $m$ and $n$ are separately done.
Compared to VMC, our approach uses quantum computers and we can, in principle, use an arbitrary unitary quantum circuit $U$ to generate a general quantum state $\ket{\psi}=U\ket{0}$ on qubits.
On the other hand, in VMC, a quantum state is supposed to have
a property that complex amplitudes $\ip{n}{\psi}$ are efficiently calculable on classical computers.
For example, when the unitary coupled-cluster ansatz~\cite{anand2021quantum} is employed for $U$, it is suggested that classical computation of $\ip{n}{\psi}$ is hard~\cite{huggins2021unbiasing}.
In that case, we may potentially have a quantum advantage in the sense that we can use wavefunctions that are difficult to be used with VMC.
We also analyze variances of expectation values obtained by the importance sampling of Eq.~\eqref{eq:ev_O_3} in the same manner as VMC in Appendix~\ref{appsec: exact sampling variance}.

Other related classical computational methods are the adaptive-sampling configuration interaction (ASCI) and similar methods~\cite{tubman2016,evangelista2014,holmes2016,schriber2016,tubman2020}, where important basis states to represent the ground state of a given Hamiltonian $H$ are adaptively chosen and the exact diagonalization within the selected basis states is iteratively performed.
From the viewpoint of searching for important basis states, our algorithm automatically picks up such basis states by projective measurements on a quantum state $\ket{\psi}$ if $\ket{\psi}$ is concentrated in the basis states, whereas ASCI relies on state-of-the-art techniques in classical computation~\cite{tubman2020} to search through a large number (possibly exponential in $N$) of basis states.
Note that our algorithm described in this article focuses on the evaluation of just the expectation value $\ev{H}{\psi}$ for a given state $\ket{\psi}$, and cannot be compared on the same footing with ASCI which performs the exact diagonalization of $H$.

\section{Summary
\label{sec:summary}}

In this study, we proposed a hybrid quantum-classical algorithm to efficiently measure expectation values of observables for quantum states concentrated in a measurement basis.
Taking the computational basis with $\ket{n}$ ($n=0,1,\cdots, 2^N -1$) as the measurement basis, we first rewrote the expectation value $\ev{O}{\psi}$ of an observable $O$ for a state $\ket{\psi}$ by expanding the state in terms of the computational basis states.
This resulted in a weighted sum of transition matrix elements $\mel{m}{O}{n}$ with interference factors $\bra{m}\ket{\psi}\bra{\psi}\ket{n}$.
We then approximated the sum with a limited number of computational basis states based on the empirical insight that quantum states in which we are interested are often concentrated: for example, in quantum chemistry, only a small number of Slater determinants such as the Hartree-Fock state are sufficient to describe low-lying electronic energy eigenstates for weakly-correlated molecules; 
though our method is applicable beyond the weak correlation.

We gave explicit procedures to measure and calculate the ingredients appearing in the truncated summation.
The single weight factors $|\ip{n}{\psi}|^2$ and the normalization factor are estimated by projective measurement for $\ket{\psi}$ in the computational basis, while the transition matrix elements $\mel{m}{O}{n}$ are efficiently calculated by classical computation.
The interference factors $\bra{m}\ket{\psi}\bra{\psi}\ket{n}$ are estimated by projective measurements of some states in the computational basis, where each of the states is realized by an appropriate unitary transformation, consisting of at most $N$ CNOT gates, applied to $\ket{\psi}$.
In our method, the number of quantities measured on quantum computers depends on the quantum state instead of the observable.
This contrasts with the conventional methods of expectation value estimation, where the observable is expanded into a plethora of Pauli strings that are separately measured on quantum computers.

For a quantitative comparison, we numerically estimated variances for estimated expectation values of Hamiltonians for electronic states of small molecules, i.e., energy expectation values.
We derived formulas to calculate the variances of expectation values estimated by our method. 
We employed the exact ground states to estimate the variances of the energy expectation values and inferred the total numbers of measurements to achieve the standard deviation of $10^{-3}$ Hartree, which is comparable to the so-called chemical accuracy in quantum chemistry.
The numerical results illustrate that our approach can outperform the conventional methods to estimate the expectation values.

As a future work, combining our method with VQE, e.g., as a subroutine to estimate expectation values, would be interesting not only for accelerating VQE, but also for envisioning how our proposal can contribute to various variational quantum algorithms.
In the numerical analysis, we used exact ground states of Hamiltonians for estimating the variances because our focus was to illustrate the performance in a single evaluation of expectation values.
In application to VQE, the estimation of expectation values is needed during the optimization procedure and hence a target quantum state may not be well concentrated, especially at initial steps of the optimization.
In such a case, the summation in our expectation value formula would require the inclusion of many basis states, possibly spoiling the effectiveness of our algorithm.
Although we focused on the application to quantum chemistry for demonstrating effectiveness of our algorithm, the algorithm is open to general applications that rely on expectation value estimation.

\begin{acknowledgements}
KM is supported by JST PRESTO Grant No.~JPMJPR2019 and JSPS KAKENHI Grant No.~20K22330.
WM is supported by JST PRESTO Grant No.~JPMJPR191A.
This work is supported by MEXT Quantum Leap Flagship Program (MEXT QLEAP) Grant No.~JPMXS0118067394 and JPMXS0120319794.
We also acknowledge support from JST COI-NEXT program Grant No.~JPMJPF2014.
A part of this work was performed for Council for Science, Technology and Innovation (CSTI), Cross-ministerial Strategic Innovation Promotion Program (SIP), ``Photonics and Quantum Technology for Society 5.0'' (Funding agency: QST).
\end{acknowledgements}

\appendix

\section{
Alternative ways to estimate interference factors
\label{appsec: interference} }

We gave an explicit algorithm to calculate the interference factors $\bra{m}\ket{\psi}\bra{\psi}\ket{n}$ or $\bra{z_r}\ket{\psi}\bra{\psi}\ket{z_{r'}}$ in Sec.~\ref{subsec:our-method}.
It was meant just for illustration, and it may be possible to calculate them in different ways hopefully with lower quantum computational costs.

For example, suppose the Hadamard gate is applied on the first qubit ($\mr{Had}_1$) of a state $\ket{\psi}$ and the projective measurement is performed for the overall state in the computational basis.
By this measurement, one can estimate the probabilities with which each of all the $N$-bit strings, from ``$0\cdots0$" to ``$1\cdots1$", appears.
When $N=4$, for instance, the probability for ``0011" is written as $|\mel{0011}{\mr{Had}_1}{\psi}|^2 = |\frac{1}{\sqrt{2}} (\bra{0011} + \bra{1011}) \ket{\psi}|^2$, contributing to the $\mathcal{A}$ term (see Eq.~\eqref{eq:intf_factor_1} for definition) of the interference factor $\bra{0011}\ket{\psi}\bra{\psi}\ket{1011}$; the probability for ``0101" is $|\mel{0101}{\mr{Had}_1}{\psi}|^2 = |\frac{1}{\sqrt{2}} (\bra{0101} + \bra{1101}) \ket{\psi}|^2$, contributing to the $\mathcal{A}$ term of $\bra{0101}\ket{\psi}\bra{\psi}\ket{1101}$.
This example indicates that one can evaluate all the $2^{N-1}$ possible $\mathcal{A}$ terms consisting of $\ket{0b_2\cdots b_N}$ and $\ket{1b_2\cdots b_N}$ ($b_2,\cdots,b_N=0,1$), solely based on a single quantum circuit.
The $\mathcal B$ term counterparts can be similarly evaluated but by measuring the state ${\mr{Had}_1}S_1\ket{\psi}$, augmented by the phase gate on the first qubit $S_1={\rm diag}(1,i)$.
Then, all the corresponding interference factors can be obtained via Eq.~\eqref{eq:intf_factor_1}, with the single-weight factors evaluated by measuring the state $\ket{\psi}$.
This procedure combined with the relation~\eqref{eq:intf_factor_2} may lead to an improved way to evaluate the interference factors which requires a smaller number of distinct quantum circuits.

We remind that, in the application to electronic structure problems with the Jordan-Wigner mapping, the above procedure gives interference factors between electronic states of {\it different electron numbers} and would not be useful in settings where the electron number is conserved.
Yet, even in such a case, similar procedures may still work to evaluate interference factors between electronic states of {\it an equal electron number} based on the parity mapping or Bravyi-Kitaev mapping~\cite{BRAVYI2002210} for encoding fermionic states onto qubits.
Or, keeping the Jordan-Wigner mapping, one can consecutively apply the CNOT and Hadamard (and phase) gates on the state $\ket{\psi}$ to obtain $\mathcal A$ ($\mathcal B$) terms with $\ket{01b_3\cdots b_N}$ and $\ket{10b_3\cdots b_N}$, both having a same electron number.
In line with these ideas, our algorithm may be better implemented with lower computational costs.

\section{
Detail of numerical analysis
\label{appsec: numerics} }

The geometries of molecules used in Sec.~\ref{sec:experiment} are summarized in Table~\ref{tab: geometries}.

\begin{table*}[] 
 \caption{Geometries of molecules. ``$(\mr{X}, (x,y,z))$" denotes three dimensional coordinates $x,y,z$ of atom X in units of \AA. 
 \label{tab: geometries}
 }
 \begin{tabular}{c|l}
 \hline \hline
 Molecule & Geometry  \\ \hline
 \ce{H2} & (H, (0, 0, 0)), (H, (0, 0, 0.735)) \\
 \ce{LiH} & (Li, (0, 0, 0)), (H, (0, 0, 1.548)) \\
 \ce{H2O} & (O, (0, 0, 0.137)), (H, (0,0.769,-0.546)), (H, (0,-0.769,-0.546)) \\
 \ce{NH3} & (N, (0,0,0.149)), (H, (0,0.947,-0.348)), (H, (0.821,-0.474,-0.348)), (H, (-0.821,-0.474,-0.348)) \\
\ce{CH4} & (C, (0, 0, 0)), (H, (0.6276, 0.6276, 0.6276)), (H, (0.6276, -0.6276, -0.6276)), \\
 & (H, (-0.6276, 0.6276, -0.6276)), (H, (-0.6276, -0.6276, 0.6276)) \\
 \ce{CO} & (C, (0, 0, 0)), (O, (0, 0, 1.128)) \\
 \ce{H2S} & (S, (0, 0, 0.1030)), (H, (0, 0.9616, -0.8239)), (H, (0, -0.9616, -0.8239)) \\
 \ce{C2H2} & (C, (0, 0, 0.6013)), (C, (0, 0, -0.6013)), (H, (0, 0, 1.6644)), (H, (0, 0, -1.6644))
\\ \hline \hline
 \end{tabular}
\end{table*}
%

\section{Approximating wavefunctions and expectation values
\label{app:fidelity}
}
In this section, we prove Eq.~\eqref{eq:bound} in the main text.

Let us consider two quantum states (density matrices) $\rho, \sigma$ and an observable $O$ which is written as $ O = \sum_i o_i E_i$ by the spectral decomposition (i.e., $o_i \in \mathbb{R}$ is an eigenvalue and $E_i$ is the corresponding projector).
We can bound the difference of expectation values of $O$ with respect to $\rho$ and $\sigma$ as
\begin{equation*}
 \begin{aligned}
 \Tr( (\rho - \sigma) O) &= \sum_i o_i \Tr((\rho - \sigma)E_i) \\
 &\leq  \sum_i |o_i| |\Tr((\rho-\sigma)E_i)| \\
 &\leq \|O\|_{\infty} \sum_i  \Tr(|\rho-\sigma|E_i) \\
 &= 2 \|O\|_{\infty} D(\rho, \sigma),
 \end{aligned}
\end{equation*}
where $\|O\|_\infty$ is the infinite norm (the largest singular value) of $O$ and $D(\rho, \sigma) = \Tr(|\rho-\sigma|)/2$ is the trace distance.
We have used an inequality $|\Tr((\rho-\sigma)E_i)| \leq \Tr(|\rho-\sigma|E_i)$ (see Section 9.2 of \cite{nielsen2010}) and the completeness relation of the projectors $\sum_i E_i = I$.
When two states are both pure, i.e., $\rho=\ket{\psi}\!\bra{\psi}$ and $\sigma=\ket{\phi}\!\bra{\phi}$, the trace distance is expressed by the fidelity $|\ip{\psi}{\phi}|^2$:
\begin{equation}
 D(\rho, \sigma) = \sqrt{1 - |\ip{\psi}{\phi}|^2}.
\end{equation}
Therefore, when we approximate an original quantum state $\ket{\psi}$ by the truncated state $\ket{\psi_R}$, the difference of expectation values is bounded by their infidelity:
\begin{equation}
 |\ev{O}{\psi} - \ev{O}{\psi_R} | \leq 2 \|O\|_{\infty} \sqrt{1 - |\ip{\psi}{\psi_R}|^2}.
\label{appeq:bound}
\end{equation}

\section{Variance formulas
\label{appsec:variance}}

In this section, we first review the formulas for variances of estimated expectation values of an observable $O$ with a given state $\ket{\psi}$ in the conventional methods based on the decomposition of $O$ into Pauli strings.
We also present the ways how to (near) optimally allocate the measurement budget to each of quantities measured in the estimation procedures in the view of minimizing
the variance, depending on whether one can access the values of the quantities to be measured prior to the measurements or not.
Then, we show the formulas for variances of estimated expectation values in our method.
The (near) optimal allocations of measurements to minimize the variance are also considered for the same situations as the conventional methods.

\subsection{
General formulas for optimal allocation of measurement budget
\label{appsubsec: general formulas for variance}
}

Both in the conventional and our methods, the variance of an expectation value can be written as
\begin{equation}
 \mr{Var}[\ev{O}{\psi}] = \sum_{\lambda} \frac{v_\lambda}{L_\lambda},
 \label{eq: general variance}
\end{equation}
where $L_\lambda$ is the number of measurements to evaluate some quantity labeled by $\lambda$ and $v_\lambda$ is the variance per single measurement for that quantity.
Minimizing the variance under the constraint of a fixed total number of measurements, $\sum_\lambda L_\lambda = L$, one obtains the optimal number of measurements for each $\lambda$ given by~\cite{wecker2015, rubin2018application, arrasmith2020operator}
\begin{equation}
 L_\lambda = L \frac{\sqrt{v_\lambda}}{\sum_{\lambda'} \sqrt{v_{\lambda'}} }.
 \label{eq: general optimal shots}
\end{equation}
Plugging this into Eq.~\eqref{eq: general variance} gives the variance for the optimal allocation of the measurement budget:
\begin{equation}
\qty( \mr{Var}[\ev{O}{\psi}])^* = \frac{\qty(\sum_{\lambda}\sqrt{v_\lambda})^2}{L}.
\label{eq: general optimal variance}
\end{equation}

This derivation tacitly assumes one can access the values of $v_\lambda$ for the state $\ket{\psi}$ of interest prior to the measurements, which would be, however, unrealistic in practical situations.
When those values are unknown, one can still make a heuristic guess for the measurement allocation to reduce the variance.
That is, one introduces a reference state $\ket{\psi^{\rm ref}}$ for which the values of $v_\lambda^{\rm ref}$ in Eq.~\eqref{eq: general variance} can be obtained by classical computation, and then optimizes the measurement allocation with respect to this state:
\begin{equation}
 L_\lambda = L \frac{\sqrt{v_\lambda^{\rm ref}}}{\sum_{\lambda'} \sqrt{v^{\rm ref}_{\lambda'}} }.
 \label{eq: general optimal shots ref}
\end{equation}
With this measurement allocation, the variance for the state $\ket{\psi}$ of interest becomes
\begin{equation}
\qty( \mr{Var}[\ev{O}{\psi}])^{\rm ref} = \frac{1}{L}\qty(\sum_{\lambda}\sqrt{v_\lambda^{\rm ref}})
\qty(\sum_{\lambda} \frac{v_\lambda}{\sqrt{v_\lambda^{\rm ref}}}).
\label{eq: general optimal variance ref}
\end{equation}

Both in the conventional and our methods for both ways of the measurement allocation, the variance as a function of the total number of measurements $L$ can be written as
\begin{equation}
 \qty( \mr{Var}[\ev{O}{\psi}])^{\rm opt}= \frac{c_v}{L},
 \label{eq: general optimal variance general}
\end{equation}
where $c_v$ is a constant consisting of $v_\lambda$ and/or $v_\lambda^{\rm ref}$ (see Eqs.~\eqref{eq: general optimal variance} and \eqref{eq: general optimal variance ref} for the definition).
We will use these general formulas when deriving the optimal allocation of measurements in the following.

\subsection{Conventional methods based on grouping of Pauli strings
\label{appsubsec: variance conventional}
}

The conventional methods for expectation value estimation utilize the expansion of the target observable $O$ by Pauli strings $P_i$, given in Eq.~\eqref{eq:O_pauli}:
\begin{align}
 O = \sum_{i=1}^M c_i P_i,
\label{appeq:ev_O_2}
\end{align}
where $c_i$ are real coefficients, and $M$ is the number of the Pauli strings.
Moreover, to save the measurement budget, one can divide Pauli strings into several groups where all Pauli strings in each group can be measured simultaneously:
\begin{align}
O &= \sum_{g} O_g, \\
O_g &= \sum_{k=1}^{M_g} c_k^{(g)} P_k^{(g)}.
\end{align}
In this expression, $g$ labels the groups; $c_k^{(g)}$ and $P_k^{(g)}$ are coefficients and Pauli strings in a group $g$, respectively; $M_g$ is the number of Pauli strings in a group $g$.
The ability to perform the simultaneous measurement is assured by imposing the condition $[P_k^{(g)}, P_{k'}^{(g)}] = P_k^{(g)} P_{k'}^{(g)} - P_{k'}^{(g)} P_k^{(g)} = 0$ in grouping, for any pair of $k,k'$ within a group $g$.
How to group Pauli strings in the original observable [Eq.~\eqref{appeq:ev_O_2}] is explained in Appendix~\ref{appsec: grouping}.

The expectation value and the (single-measurement) variance for $O_g$ are given by 
\begin{align}
 \mr{Exp}[O_g] &= \ev{O_g}{\psi}, \label{eq: group exp} \\
 \mr{Var}[O_g]  &= \ev{O_g^2}{\psi} - \ev{O_g}{\psi}^2,
 \label{eq: group var}
\end{align}
respectively.
When one performs $L_g$ times of measurements for each group $g$, the variance for the overall expectation value $\ev{O}{\psi}$ is given by
\begin{equation}
 \mr{Var}[\ev{O}{\psi}]_{\mr{grouping}} = \sum_g \frac{\mr{Var}[O_g]}{L_g},
\end{equation}
because measurements for different groups are independent.

When one can access the 
value of $\mr{Var}[O_g]$, or $\ev{O_g}{\psi}$ and $\ev{O_g^2}{\psi}$, it is possible to optimize the number of measurements $L_g$ for each group $g$ by using Eq.~\eqref{eq: general optimal shots} and the minimal variance under the constraint $\sum_g L_g = L$ is obtained as
\begin{widetext}
 \begin{align}
 \qty(\mr{Var}[\ev{O}{\psi}]_{\mr{grouping}})^*  & = \frac{1}{L} \qty(\sum_g \sqrt{\mr{Var}[O_g]})^2 \notag \\
 &= \frac{1}{L} \qty( \sum_g \sqrt{ \sum_{k,k'} c_k^{(g)} c_{k'}^{(g)} \qty( \ev{P_k^{(g)} P_{k'}^{(g)}}{\psi} - \ev{P_k^{(g)}}{\psi} \! \ev{P_{k'}^{(g)}}{\psi})  })^2,
 \label{eq: best variance conventional fci}
\end{align}
\end{widetext}
according to Eq.~\eqref{eq: general optimal variance}.
However, in most practical cases, one does not know the 
values for $\ev{O_g^2}{\psi}$ and $\ev{O_g}{\psi}$.
In such cases, the optimal numbers of measurements can be heuristically guessed by using expectation values for Haar random states~\cite{wecker2015, 
gokhale2020ON3}.
The expectation values for Haar random state satisfy $\ev{P_i}_{\mr{Haar}} = 0$ for any Pauli string $P_i$, and the guessed variance is
\begin{equation}
 \mr{Var}[O_g]^{\mr{Haar}} = \sum_{k} (c_k^{(g)})^2.
\end{equation}
By employing the numbers of measurements Eq.~\eqref{eq: general optimal shots} for these variances and putting them into Eq.~\eqref{eq: general variance}, the resulting variance for this heuristic choice of the numbers of measurements is
\begin{align}
 &\qty(\mr{Var}[\ev{O}{\psi}]_{\mr{grouping}})^{\mr{Haar}} \notag \\
 &= \frac{1}{L} \qty( \sum_g \frac{\mr{Var}[O_g]}{ \sqrt{\mr{Var}[O_g]^{\mr{Haar}}} } ) \qty(\sum_g \sqrt{\mr{Var}[O_g]^{\mr{Haar}}} ).
 \label{eq: best variance conventional haar}
\end{align}
This corresponds to  the general formula of Eq.~\eqref{eq: general optimal variance ref}.
Here we adopt the Haar random state to decide the measurement allocation in part because of its simplicity, but the use of other classically tractable states, e.g., the state obtained by CISD method, is also possible~\cite{gonthier2020identifying}.
The latter could in principle reduce the variance of the expectation value for some state of interest, but this is beyond the scope of this work.

\subsection{Our method
\label{appsubsec: variance ours}
}

Here we give the formulas for the variance of the expectation value $\ev{O}{\psi}$ in our method as well as the (near) optimal allocation of the measurement budget to minimize the variance of the estimated expectation value.

\subsubsection{Variance of $\ev{O}{\psi}$}

We estimate the expectation value by the algorithm described in Sec.~\ref{subsec:our-method}.
The expectation value is approximated as
\begin{align}
\ev{O}{\psi}
\approx {\mc{N}_R}^2 \sum_{r,r'=1}^R |\ip{z_r}{\psi}|^2 |\ip{z_{r'}}{\psi}|^2
\frac{\mel{z_r}{O}{z_{r'}}}{
\ip{z_r}{\psi} \ip{\psi}{z_{r'}}
},
\end{align}
which is evaluated by the measured quantities $f_r, c_R, A_r$, and $B_r$:
\begin{align}
|\ip{z_r}{\psi}|^2 &\simeq f_r , \\
\mc{N}_R &\simeq c_R , \\
\bra{z_1}\ket{\psi}\bra{\psi}\ket{z_r}
 &\simeq g_r \quad (r=2,\cdots, R),\\
\bra{z_r}\ket{\psi}\bra{\psi}\ket{z_1}
 &\simeq g_r^* \quad (r=2,\cdots, R),\\
\bra{z_r}\ket{\psi}\bra{\psi}\ket{z_{r'}}
 &\simeq \frac{g_r^* g_{r'} }{f_1}
 \: (r\neq r' \text{and} \: r,r' = 2,\cdots, R),
\end{align}
with
\begin{align}
g_r = A_r + iB_r -\frac{1+i}{2}(f_1 + f_r).
\end{align}
In the following derivation of the variance formula, we omit the effect of normalization, $\mc{N}_R$, to ease the analysis of the variance.
More precisely, we approximate $\ket{\psi}$ by  $\ket{\psi_R}$ and set $\mc{N}_R=1$ throughout in this subsection.
Numerical results in Sec.~\ref{subsec: shot experiment} are computed under this assumption.
Since we consider $\mc{N}_R \approx 1 - 10^{-4}$ in our study, the error in the variance estimations caused by this approximation will be small and we believe it does not affect the conclusion in Sec.~\ref{subsec: shot experiment}.
This point is explicitly examined in Appendix~\ref{appsec: normalitzaion effect} and we find the error is small.

The expectation value $\ev{O}{\psi}$ depends on measured quantities $\{f_r\}_{r=1}^R, \{A_r\}_{r=2}^R$, and $\{B_r\}_{r=2}^R$.
The variance of $\ev{O}{\psi}$ is given by the error propagation formula as
\begin{align}
& \mr{Var}[\ev{O}{\psi}]_\text{our method} \notag \\
& = 
 \sum_{r=1}^R \sum_{r'=1}^R
 \frac{\partial \ev{O}}{\partial f_r} \frac{\partial \ev{O}}{\partial f_{r'}}{\rm Cov}[f_r, f_{r'}] \notag\\
 &\quad +\sum_{r=2}^R \left( \frac{\partial \ev{O}}{\partial A_r} \right)^2 {\rm Var}[A_r]
 +\sum_{r=2}^R \left( \frac{\partial \ev{O}}{\partial B_r} \right)^2
 {\rm Var}[B_r],
\label{eq:error-prop}
\end{align}
where we denote $\ev{O} = \ev{O}{\psi}$ and the derivatives are explicitly given by
\begin{align}
&\frac{\partial \ev{O}}{\partial f_1} \notag \\
&= \mel{z_1}{O}{z_{1}}
   +2{\rm Re}\sum_{r=2}^R
     \frac{f_r}{g_r} \left[  1+\left(\frac{1+i}{2}\right)\frac{f_1}{g_r} \right]
   \mel{z_1}{O}{z_{r}} \notag\\
   &\quad +2{\rm Re}\sum_{r=2}^R \sum_{r'=r+1}^R
     \frac{f_r f_{r'}}{g_r^* g_{r'}} \notag\\
   &\quad \times \left[ 1+\left(\frac{1+i}{2}\right)\frac{f_1}{g_{r'}}
     +\left(\frac{1-i}{2}\right)\frac{f_1}{g_{r}^*} \right]
   \mel{z_r}{O}{z_{r'}},
\label{eq:pdv_O_f1}
\end{align}
and 
\begin{align}
&\frac{\partial \ev{O}}{\partial f_r}\notag\\
&= \mel{z_r}{O}{z_r}
   +2 {\rm Re} \frac{f_1}{g_r}
     \left[  1+\left(\frac{1+i}{2}\right)\frac{f_r}{g_r} \right]
   \mel{z_1}{O}{z_r} \notag\\
   &\quad +2{\rm Re}\sum_{r' \geq 2, \neq r}
     \frac{f_1 f_{r'}}{g_r^* g_{r'}}
     \left[ 1+\left(\frac{1-i}{2}\right) \frac{f_r}{g_{r}^*} \right]
    \mel{z_r}{O}{z_{r'}},\\
&\frac{\partial \ev{O}}{\partial A_r} \notag\\
&= -2{\rm Re} \frac{f_1 f_r}{g_r^2} \pqty{  \mel{z_1}{O}{z_{r}}
    +\sum_{r'\geq 2, \neq r}
     \frac{f_{r'}}{g_{r'}^*} \mel{z_{r'}}{O}{z_{r}} }, \\
&\frac{\partial \ev{O}}{\partial B_r} \notag\\
&= 2{\rm Im} \frac{f_1 f_r}{g_r^2}
   \pqty{
   \mel{z_1}{O}{z_{r}}
   +\sum_{r'\geq 2, \neq r}
     \frac{f_{r'}}{g_{r'}^*} \mel{z_{r'}}{O}{z_{r}}
     },
\label{eq:pdv_O_Br}
\end{align}
for $r=2,\cdots,R$, and $\mr{Cov}[f_r, f_{r'}], \mr{Var}[A_r]$, and $\mr{Var}[B_r]$ represent the covariance between $f_r$ and $f_{r'}$, the variance of $A_r$, and the variance of $B_r$, respectively.
Note that we evaluate all $f_1, \cdots, f_R$ at the same time (in one experiment) and the covariance between $f_r$ and $f_{r'}$ is non-zero.

As described in Sec.~\ref{sec:method}, the values of measured $f_1, \cdots, f_R$ are evaluated by the numbers of occurrence of the result $z_1, \cdots, z_R$ in the projective measurement for $\ket{\psi}$ which is repeated $L_f$ times.
The numbers of the occurrence $T_1, \cdots, T_R$ obey the multinomial distribution, and the variance and covariance are given by
\begin{align}
\mr{Var}[T_r] &= L_f p_r(1-p_r),\\
\mr{Cov}[T_r, T_{r'}] &= - L_f p_r p_{r'} \quad (r \neq r'),
\end{align}
where $p_r = |\ip{z_r}{\psi}|^2$ is the probability for measuring $z_r$ (note that only $z_1,\cdots,z_R$ are observed since we approximate $\ket{\psi}$ by $\ket{\psi_R}$).
Since $f_r = T_r /L_f$, we reach
\begin{align}
\mr{Var}[f_r] &= \frac{f_r(1-f_r)}{L_f} \quad (r=1,\cdots, R), \label{eq: variance of f} \\
\mr{Cov}[f_r, f_{r'}]
& = -\frac{f_r f_{r'}}{L_f} \quad (r\neq r').
\label{eq: covariance of f}
\end{align}
We define $\mr{Cov}[f_r, f_r] = \mr{Var}[f_r]$ for simplifying the notation.
On the other hand, $A_r$ ($r=2,\cdots,R$) is measured as a frequency of obtaining the result ``0" when measuring a state $U_{z_1, z_{r}}\ket{\psi}$ in the computational basis.
Since we repeat the measurement $L_{A_r}$ times for each $r$, the variance of $A_r$ is given by a simple binomial distribution of the probability $|\mel{0}{U_{z_1,z_r}}{\psi}|^2 \simeq A_r$:
\begin{align}
\mr{Var}[A_r] = \frac{A_r(1-A_r)}{L_{A_r}},
\label{eq: variance of A}
\end{align}
for $r = 2,\cdots, R$.
Similarly, the variance of $B_r$ for $L_{B_r}$ measurements is given by
\begin{align}
\mr{Var}[B_r] = \frac{B_r(1-B_r)}{L_{B_r}},
\label{eq: variance of B}
\end{align}
for $r = 2,\cdots, R$.

By putting Eqs.~\eqref{eq: variance of f}-\eqref{eq: variance of B} into Eq.~\eqref{eq:error-prop}, we 
eventually reach the formula in the form of Eq.~\eqref{eq: general variance}:
\begin{align}
\mr{Var}[\ev{O}{\psi}]_\text{our method}
= \frac{v_f}{L_f}
+\sum_{r=2}^R \pqty{
 \frac{v_{A_r}}{L_{A_r}} +\frac{v_{B_r}}{L_{B_r}}},
\label{eq: variance of our method}
\end{align}
where
\begin{align}
v_f 
&= \sum_{r=1}^R
 \left( \frac{\partial \ev{O}}{\partial f_r} \right)^2 f_r (1-f_r) 
 -\sum_{r\neq r'}
 \frac{\partial \ev{O}}{\partial f_r} \frac{\partial \ev{O}}{\partial f_{r'}} f_r f_{r'},\notag\\
v_{A_r}
&= \left( \frac{\partial \ev{O}}{\partial A_r} \right)^2 A_r (1-A_r),\notag\\ 
v_{B_r}
&= \left( \frac{\partial \ev{O}}{\partial B_r} \right)^2 B_r (1-B_r).
\label{eq: variance of our method_where}
\end{align}

\subsubsection{
Optimal allocation of measurement budget
}

We discuss the optimization of the numbers of measurements $L_f, \{L_{A_r}\}_{r=2}^R, \{ L_{B_r}\}_{r=2}^R$ to minimize the variance~\eqref{eq: variance of our method}.
We fix the total number of measurements:
\begin{align}
L = L_f +\sum_{r=2}^R \pqty{L_{A_r} +L_{B_r} }.
\label{eq:tot}
\end{align}
We again consider the situation where one can use information of values for $f_r, A_r, B_r$, or the situation where one cannot.

When the values can be used, by the results in Sec.~\ref{appsubsec: general formulas for variance}, the optimal numbers of measurements are
\begin{align}
L_f = L \sqrt{\frac{v_f}{v}},~
L_{A_r} = L\sqrt{\frac{v_{A_r}}{v}},~
L_{B_r} = L\sqrt{\frac{v_{B_r}}{v}},
\label{eq: shots for best variance ours}
\end{align}
where $ \sqrt{v} = 
 \sqrt{v_f} +\sum_{r=2}^R \pqty{\sqrt{v_{A_r}} +\sqrt{v_{B_r}}}$.
The minimal variance achieved by those numbers of measurements is
\begin{align}
& \qty(\mr{Var}[\ev{O}{\psi}]_{\text{our method}})^* \notag \\
&= \frac{1}{L}
\pqty{\sqrt{v_f} +\sum_{r=2}^R \pqty{\sqrt{v_{A_r}} +\sqrt{v_{B_r}} } }^2.
\label{eq: best variance ours}
\end{align}

When one cannot utilize the values of $f_r, A_r, B_r$ prior to the measurements, we propose using the following heuristic wavefunction,
\begin{equation}
 \ket{\psi_w} = \sqrt{w} \ket{z_1} + \sum_{r=2}^R \sqrt{\frac{1-w}{R-1}}\ket{z_r},
 \label{eq: heuristic wfn}
\end{equation}
where $w$ is a real parameter in $0<w<1$, set by hand.
This state mimics a quantum state concentrating especially on $\ket{z_1}$ with a (large) weight $w$ and having equal weights for the other computational basis states $\ket{z_2}, \cdots, \ket{z_R}$. 
For the ground states of molecular Hamiltonians, the Hartree-Fock states (mean-field solutions of the Hamiltonians) typically have large overlaps with the exact ground states; so the use of the heuristic state $\ket{\psi_w}$ might be helpful to guess the variances of $f_r, A_r, B_r$ and to choose the numbers of measurements leading to a relatively small variance of the expectation value $\ev{O}{\psi}$. 
We use the values of $f_r, A_r, B_r$ for this state,
\begin{align}
 f_1^{w} &= w, \quad f_2^w = \cdots = f_R^w = \frac{1-w}{R-1}, \\
 A_2^{w} &= \cdots = A_R^{w} = \frac{1}{2} \abs{\sqrt{w} + \sqrt{\frac{1-w}{R-1}} }^2, \\
 B_2^{w} &= \cdots = B_R^{w} = \frac{1}{2} \abs{\sqrt{w} + i\sqrt{\frac{1-w}{R-1}} }^2,
\end{align}
and guess the coefficients in Eq.~\eqref{eq: variance of our method} as
\begin{align}
v_f^w
&= \sum_{r=1}^R
 \left. \qty(\pdv{\ev{O}}{f_r})^2 \right|_{\ket{\psi_w}} f_r^w (1-f_r^w) \notag \\
& -\sum_{r\neq r'}  \left. \pdv{\ev{O}}{f_r} \right|_{\ket{\psi_w}} \left. \pdv{\ev{O}}{f_{r'}} \right|_{\ket{\psi_w}} f_r^w f_{r'}^w,\notag\\
v_{A_r}^w
&= \left. \left( \frac{\partial \ev{O}}{\partial A_r} \right)^2 \right|_{\ket{\psi_w}} A_r^w (1-A_r^w),\notag\\ 
v_{B_r}^w
&= \left. \left( \frac{\partial \ev{O}}{\partial B_r} \right)^2 \right|_{\ket{\psi_w}} B_r^w (1-B_r^w),
\end{align}
where $\left. \pdv{\ev{O}}{f_r}\right|_{\ket{\psi_w}}, \left. \pdv{\ev{O}}{A_r}\right|_{\ket{\psi_w}}, \left. \pdv{\ev{O}}{B_r}\right|_{\ket{\psi_w}}$ mean that we use $f_r^w, A_r^w$, and $B_r^w$ in the formulas Eqs.~\eqref{eq:pdv_O_f1}-\eqref{eq:pdv_O_Br}.
We decide the measurement allocation for the state $\ket{\psi}$, with which the expectation value is estimated, based on these guessed variances,
\begin{align}
L_f^w = L\sqrt{\frac{v_f^w}{v^w}},~
L_{A_r}^w = L\sqrt{\frac{v_{A_r}^w}{v^w}}~,
L_{B_r}^w = L\sqrt{\frac{v_{B_r}^w}{v^w}},
\label{eq: shots for heuristic variance ours}
\end{align}
where $ \sqrt{v^w} = \sqrt{v_f^w} +\sum_{r=2}^R \pqty{\sqrt{v_{A_r}^w} +\sqrt{v_{B_r}^w}}$.
The variance with this measurement allocation is
\begin{align}
& \qty(\mr{Var}[\ev{O}{\psi}]_{\text{our method}})^{w} \notag \\
&= \frac{1}{L}
\pqty{\sqrt{v_f^w} +\sum_{r=2}^R \pqty{\sqrt{v_{A_r}^w} +\sqrt{v_{B_r}^w} } }  \notag \\
& \quad \times \pqty{ \frac{v_f}{\sqrt{v_f^w}} +\sum_{r=2}^R \pqty{ \frac{v_{A_r}}{\sqrt{v_{A_r}^w}} + \frac{v_{B_r}}{\sqrt{v_{B_r}^w}} } }.
\label{eq: heuristic variance ours}
\end{align}
This corresponds to  the general formula of Eq.~\eqref{eq: general optimal variance ref}.

\section{Grouping of Pauli strings
\label{appsec: grouping} }
In this section, we review algorithms to divide Pauli strings contained in an observable into simultaneously-measurable groups, which are used in numerical analysis in Sec.~\ref{sec:experiment} to estimate the variances of the conventional methods.

As explained in the main text and Appendix ~\ref{appsubsec: variance conventional}, the grouping of Pauli strings appearing in a Hamiltonian can reduce the variance of the estimated expectation value~\cite{crawford2021efficient}.
We define two types of commutability among Pauli strings $P = \otimes_{s=1}^N P^{[s]}$, where $P^{[s]} \in \{I, X, Y, Z\}$ is the Pauli operator acting on a qubit $s=1,\cdots,N$:
one type is {\it qubit-wise commuting}~\cite{kandala2017, verteletskyi2020} and the other is {\it generally commuting}.
Two Pauli strings $P = \otimes_{s=1}^N P^{[s]}$ and $Q = \otimes_{s=1}^N Q^{[s]}$ are qubit-wise commuting if and only if $[P^{[s]}, Q^{[s]}] = 0$ for all $s=1,\cdots,N$, while two Pauli strings are generally commuting if and only if $[P, Q]=0$.
It is evident that if $P$ and $Q$ are qubit-wise commuting then they are also generally commuting.

For a group of Pauli strings $\{ P_1, P_2, .., P_{M_g} \}$ in which all elements are qubit-wise commuting each other, one can construct a quantum circuit $U_g$ consisting of at most $N$ one-qubit gates that diagonalizes all elements of the group simultaneously, i.e., $U_g^\dag P_i U_g =$ (a sum of Pauli strings consisting only from $I$ and $Z$) for all $i=1,\cdots,M_g$.
In other words, by using $U_g$, one can perform the projective measurement of $O_g = \sum_{k=1}^{M_g} c_k^{(g)} P_k^{(g)}$ whose expectation value and variance are given in Eqs.~\eqref{eq: group exp} and \eqref{eq: group var}, respectively.
When a group of Pauli strings in which all elements are generally commuting each other, it is possible to construct a quantum circuit $V_g$ consisting of $\mathcal{O}(N^2 / \log N)$ two-qubit gates and perform a projective measurement of $O_g$~\cite{yen2020,aaronson2004improved}.

In the numerical analysis in Sec.~\ref{sec:experiment}, we employ the {\it sorted insertion} algorithm~\cite{crawford2021efficient} to determine the grouping of Pauli stings contained in the Hamiltonian $H=\sum_i c_i P_i$ with $P_i \neq I^{\otimes N}$.
This algorithm works in reasonable classical computational time and it has been shown that the resulting groups exhibit smaller measurement cost than those obtained by other groping strategies~\cite{crawford2021efficient}.
We first sort Pauli strings in descending order of the magnitudes of the coefficients $|c_i|$.
We then pick up the Pauli string which has the largest magnitude, say $|c_L|$, and create the first group $G_1 = \{P_L\}$.
We grow the group $G_1$ by 
going through the sorted Pauli strings in descending order and 
putting a Pauli string in 
the group if it is {\it commuting} with all other elements already included in $G_1$.
Here, {\it commuting} means either 
qubit-wise commuting or generally commuting, depending on which we want to use in the numerical analysis.
The growth of $G_1$ is terminated once all the Pauli strings are exhausted, and we then pick up the Pauli string with the largest magnitude among the remaining Pauli strings, i.e., not included in $G_1$, and make a new group $G_2$.
We again grow $G_2$ in the same manner as we do for $G_1$ and repeat this procedure until all the Pauli strings contained in the Hamiltonian are included in one of the groups, $G_1, \cdots, G_n$.

\section{Effect of normalization factor in estimating variance
\label{appsec: normalitzaion effect}}

In the analysis of the variance in Appendix~\ref{appsubsec: variance ours} and the numerical analysis in Sec.~\ref{subsec: shot experiment},
we ignore the effect of the normalization factor $\mc{N}_R$ on the variance of estimated expectation values.
For completeness, we perform additional numerical calculation which includes the effect of the normalization factor and more closely resembles an actual experimental setup on real quantum computers.

To determine the variance per single measurement of our method in a realistic setup using sampling, we run the following calculations.

\begin{itemize}
 \item Procedure 1:
   \begin{itemize}
\item Step 1: Choose and fix an integer $L_f$. Perform the projective measurement on $\ket{\psi}$ in the computational basis 
$L_f$ times and pick up the $\tilde{R}$-most-frequent basis states, $\ket{\tilde{z}_1},\cdots, \ket{\tilde{z}_{\tilde{R}}}$. The value of $\tilde{R}$ is chosen as the smallest integer $R'$ such that the sum of the relative frequencies of the $R'$-most-frequent basis states becomes larger than $1-10^{-4}$. The relative frequency of occurrence of each basis is used as the estimate of $f_r$ in Step 3.
The normalization factor $\mc{N}_{\tilde{R}}$ is also estimated 
by $1/\mc{N}_{\tilde{R}}^2 \simeq \sum_{r=1}^{\tilde{R}} f_r$.
\item Step 2: For the basis states selected in the previous step, classically calculate the values of $f_r \: (r=1,\cdots,\tilde{R})$ and $A_r, B_r \: (r=2,\cdots,\tilde{R})$ 
using the state $\ket{\psi}$ (not $\ket{\psi_{\tilde{R}}}$).
Determine the numbers of measurements $L_{A_r}$ ($L_{B_r}$) for $A_r$ ($B_r$) using the formulas~\eqref{eq:tot} and \eqref{eq: shots for best variance ours}.
Note that we round the numerical values of $L_{A_r}$ and $L_{B_r}$ to nearest integers and that the total number of measurements, $L = L_f + \sum_{r=2}^{\tilde{R}}(L_{A_r} + L_{B_r})$, depends on the basis states  $\ket{\tilde{z}_1},\cdots, \ket{\tilde{z}_{\tilde{R}}}$ selected in Step 1 because we fix $L_f$.
\item Step 3: Perform the measurement $L_{A_r}$ ($L_{B_r}$) times to estimate $A_r$ ($B_r$) for each $r$ and obtain estimated values of them.
Combine the results with the estimates of $f_r$ and $\mc{N}_R$ in Step 1 to calculate an estimate of the energy expectation value by using Eq.~\eqref{eq: ev_approx}.
    \end{itemize}
\item Procedure 2: Repeat Procedure 1, $M$ times, and get pairs of an estimated energy and the total number of measurements $(E_1, L_1), \cdots, (E_M, L_M)$.
Calculate the sample mean $\mu$ and unbiased sample standard deviation $\sigma'$ of $E_1, \cdots, E_M$.
Calculate the average of the total numbers of measurements $\bar{L} = (L_1 + \cdots L_M)/M$ to obtain an approximate standard deviation of the energy expectation value per single measurement, $\sigma_1 \equiv \sigma' \times \sqrt{\bar{L}}$.
\item Procedure 3: Repeat Procedure 2 for $M'$ times and get pairs of the mean of the estimated energy expectation value and the standard deviation per single measurement, $(\mu_1, (\sigma_1)_1), \cdots, (\mu_{M'}, (\sigma_1)_{M'})$.
Take the sample mean and unbiased sample standard deviation of $\{\mu_i\}_{i=1}^{M'}$ and $\{ (\sigma_1)_i\}_{i=1}^{M'}$.
\end{itemize}
We perform calculation of Procedure 3 for \ce{LiH} and \ce{H2O} molecules with $L_f =10^5, M=100, M'=100$.
(The numerical setup is the same as in Sec.~\ref{sec:experiment}.)
We denote the sample means of $\{\mu_i\}_{i=1}^{M'}$ and $\{ (\sigma_1)_i\}_{i=1}^{M'}$ by $E_\mr{sampled}$ and $\sigma_\mr{1shot,sampled}$, respectively.
We compare $E_\mr{sampled}$ and $\sigma_\mr{1shot,sampled}$ with the values $E_R$ and $\sigma_\mr{1shot, R}$ obtained 
by ignoring the normalization factor, which is presented in the main text.
Note $\sigma_\mr{1shot, R}$ is the standard deviation of $\ev{H}{\psi}$ analytically evaluated for $L=1$.

The results are summarized in Table~\ref{apptab: normalization}.
The {\it sampled} version of the estimated energy is slightly larger than that obtained by ignoring the normalization factor, but 
the deviation from the exact energy is as small as $\mc{O}(10^{-4})$ Hartree.
As for the standard deviation per single measurement, $\sigma_\mr{1shot,sampled}$ is also slightly larger than $\sigma_\mr{1shot,R}$ for each of molecules, but the difference is within the uncertainty and would not give any significant changes to the results in Fig.~\ref{fig:shot}.
These results numerically support the numerical analysis in the main text and the analysis in Appendix~\ref{appsubsec: variance ours}.

\begin{table*}[]
\caption{Effect of normalization. All units are Hartree.
\label{apptab: normalization}}
\begin{tabular}{c|cccc}
\hline \hline
 molecule &  $E_R - E_\mr{exact}$ &  $E_{\mr{sampled}} - E_{\mr{exact}}$ & $\sigma_\mr{1shot, R}$ & $\sigma_\mr{1shot, sampled}$ \\ \hline 
 \ce{LiH} & $2.4 \times 10^{-4}$ & $ (2.7 \pm 0.8) \times 10^{-4}$ & $0.429$ & $0.435 \pm 0.03$ \\ 
 \ce{H2O} & $2.8 \times 10^{-4}$ & $(4.0 \pm 3.5) \times 10^{-4}$ & $1.77$ & $1.88 \pm 0.14$
 \\ \hline \hline
\end{tabular}
\label{tab: sampling experiment}
\end{table*}

\section{Numerical analysis for the nitrogen molecule with various bond lengths
\label{appsec: N2 result}}

\begin{figure}
\includegraphics[width=0.45\textwidth]{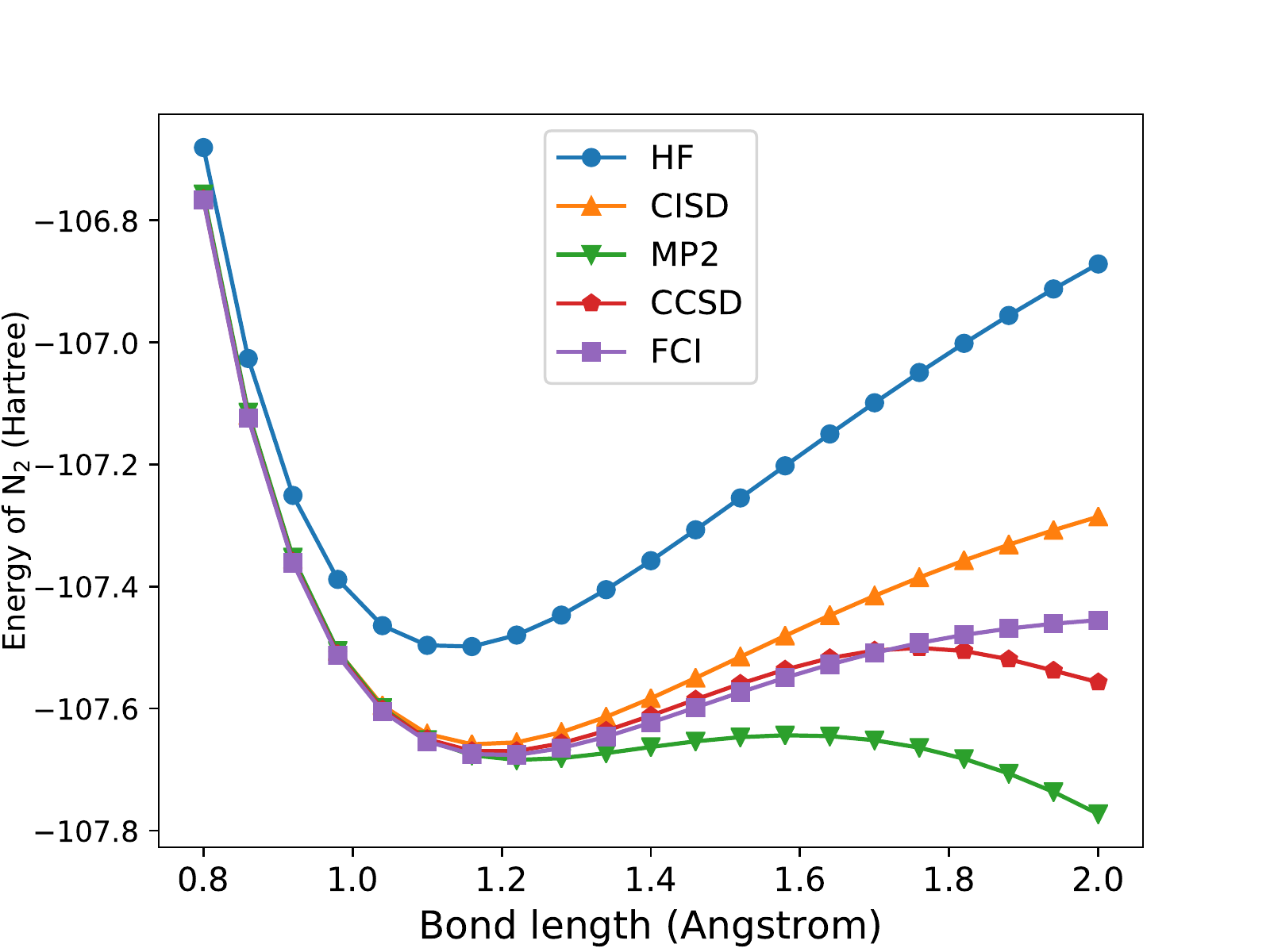}
\caption{
Potential energy curves for the ground state of \ce{N2} calculated by various classical computational methods.
The ground state energies are calculated by the Hartree-Fock (HF) method, configuration interaction singles and doubles (CISD), second-order M{\o}ller–Plesset perturbation theory (MP2), coupled cluster singles and doubles (CCSD), and full configuration interaction (FCI, or exact ground state energy).}
\label{fig: N2 PEC}
\end{figure}

\begin{table}
\caption{
Properties of the Hamiltonian and the ground state for \ce{N2} shown for various values of the bond length. 
``Pauli terms" is the number of Pauli strings contained in the Hamiltonian, $R$ is the number of computational basis states retained in the approximate expectation value (Eq.~\eqref{eq: ev_approx}) for the exact ground state, and $E_R- E_{\rm exact}$ is the difference between the energy expectation values for the truncated state and the exact ground state. 
}
\begin{center}
\begin{tabular}{c|cccccc}
\hline\hline
Bond length [\AA] & Pauli terms & $R$ & $E_R - E_{\rm exact}$ [$10^{-4}$ Hartree] \\
\hline
1.0 & $2239$ & $138$ & $7.0$ \\
1.2 & $2239$ & $183$ & $5.5$ \\
1.6 & $2239$ & $237$ & $3.3$ \\
2.0 & $2239$ & $209$ & $1.7$  \\
\hline\hline
\end{tabular}
\end{center}
\label{tab:N2}
\end{table}

\begin{figure*}
\includegraphics[width=0.49\textwidth]{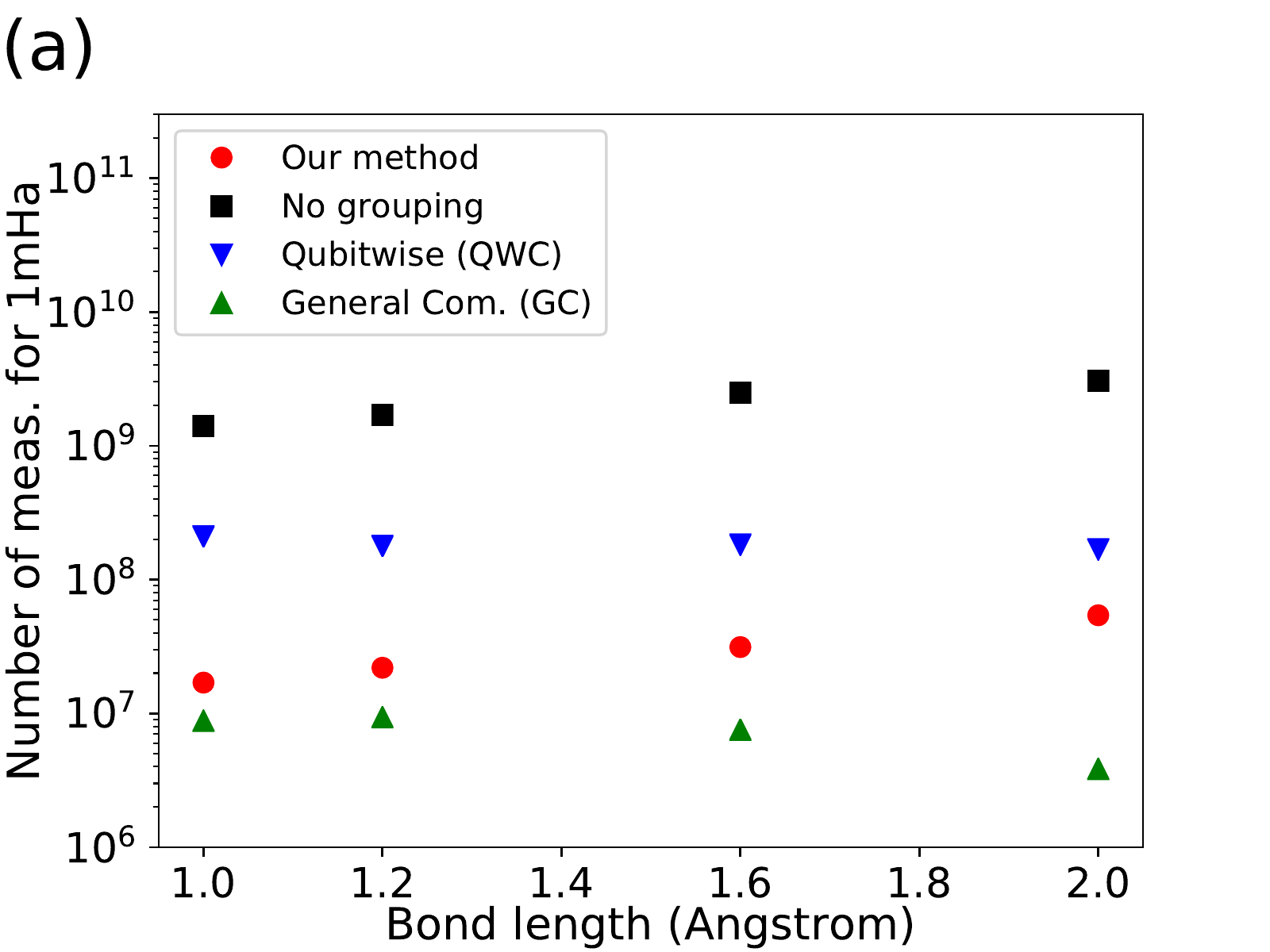} 
\includegraphics[width=0.49\textwidth]{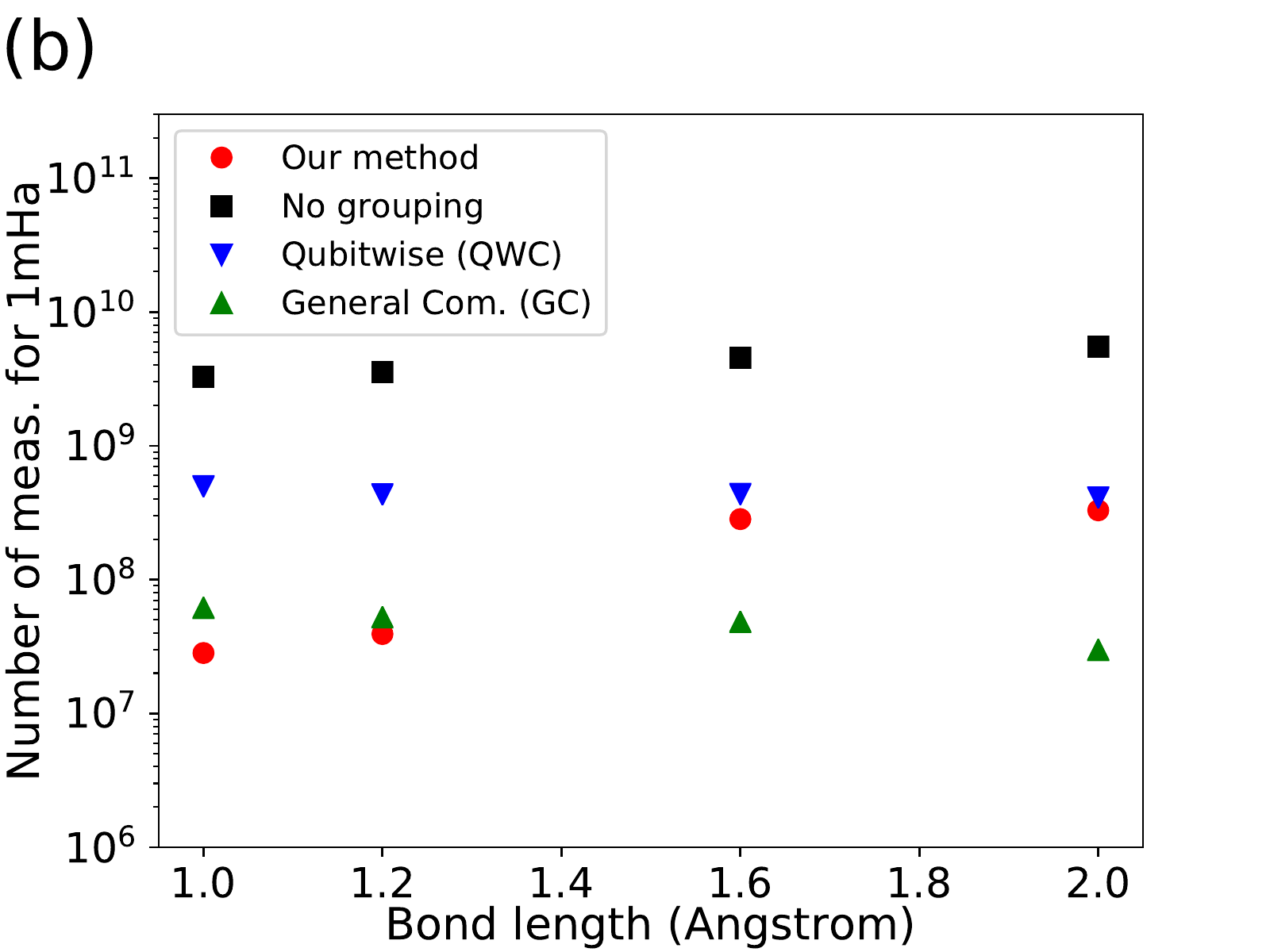} 
\caption{
The total number of measurements required to achieve the standard deviation of $10^{-3}$ Hartree in the energy expectation value estimation with the ground state for the nitrogen molecule \ce{N2} with various bond lengths, based on a different estimation method and measurement allocation strategy.
(a) The results based on the measurement allocations optimized for the exact ground states (obtained by FCI calculations).
(b) The results based on the heuristic measurement allocations.
}
\label{fig:N2 shot}
\end{figure*}

In this section, we perform the same analysis as in Sec.~\ref{sec:experiment} for the nitrogen molecule \ce{N2} with various bond lengths.
\ce{N2} with a stretched geometry is known as an example for a correlated electron system where the theoretical descriptions based on a single reference state (typically Hartree-Fock state) fail.
We investigate whether our method based on the approximation of the wavefunction by a limited number of computational basis states or Slater determinants still works in such a case.

First, we calculate the potential energy curves for the ground state of \ce{N2} by various classical computational methods with the STO-3G minimal basis set (Fig.~\ref{fig: N2 PEC}).
As the bond length becomes larger than the equilibrium distance around 1.2\AA, the energies computed by the classical methods based on the single-reference state (HF, CISD, MP2, CCSD) deviate more and more from the exact one.
This may indicate that the ground state of stretched \ce{N2} is not well-described by a small number of Slater determinants, casting doubt on the assumption of concentration of the state in our approach.
But, there may be a possibility that the ground state is still concentrated enough even though the degree of concentration is beyond the capability of those classical methods.

Next, we perform the same analysis as in Sec.~\ref{subsec: wfn analysis} by choosing four different bond lengths: 1.0\AA, 1.2\AA, 1.6\AA, and 2.0\AA.
In Table~\ref{tab:N2} (corresponding to Table~\ref{tab:info}), the number of the Pauli strings in the Hamiltonian and the value of $R$, which is the minimum number of computational basis states satisfying $1 - \abs{\braket{\psi_R}{\psi}}^2 \leq 10^{-4}$, are shown for each bond length.
The value of $R$ slightly increases when the bond length becomes larger than 1.2\AA, which is close to the equilibrium distance, possibly reflecting the nature of the wavefunction that is not described by the single-reference methods.

Finally, we evaluate the standard deviation in the estimation of energy expectation value for the ground state of \ce{N2} in the same way as presented in Sec.~\ref{subsec: shot experiment} (all the conditions for numerical calculations are taken to be the same).
Then, in Fig.~\ref{fig:N2 shot}, the estimation for the total number of measurements to realize the standard deviation of $10^{-3}$ Hartree is shown for each of the bond lengths and for various methods (presented in the similar way as in Fig.~\ref{fig:shot}).
Our method gets slightly worse as the bond length increases, but is still more efficient than QWC.
The overall tendency of the results between the different methods (no grouping, QWC or general-commuting grouping, and our method) is the same as discussed in Sec.~\ref{subsec: shot experiment}.

The results presented in this section 
demonstrate a case where our method works efficiently even for a correlated electron system that cannot be well-described by the single-reference methods but still has a concentrated wavefunction.
This supports the applicability of our method beyond weakly-correlated molecules. 

\section{Variance for explicit importance sampling
\label{appsec: exact sampling variance}}
We point out that the variance of expectation values evaluated by the explicit importance sampling of Eq.~\eqref{eq:ev_O_3},
\begin{equation}
\ev{O}{\psi} = {\sum_{m,n}}^{\prime}
    \abs{\bra{m}\ket{\psi}}^2\abs{\bra{n}\ket{\psi}}^2
    \frac{\mel{m}{O}{n} }{\bra{m}\ket{\psi} \bra{\psi}\ket{n}},
\end{equation}
may be exponentially large in the number of qubits $N$. 
In the discussion here, we assume the quantum state $\ket{\psi}$ is real for simplicity.
Here we mean by the explicit importance sampling that one samples a quantity $f(m,n) = \frac{\mel{m}{O}{n} }{\bra{m}\ket{\psi} \bra{\psi}\ket{n}}$ with a probability of $p(m,n) = \abs{\bra{m}\ket{\psi}}^2\abs{\bra{n}\ket{\psi}}^2$ and takes an average of the results from repeated sampling.
If we could evaluate $f(m, n)$ exactly without any sampling error, the variance of the sampling would be
\begin{align*}
 &{\sum_{m,n}}' (f(m,n))^2 p(m,n) - \qty({\sum_{m,n}}'f(m,n) p(m,n))^2 \\
 =& {\sum_{m,n}}' \frac{\mel{m}{O}{n}^2 }{\bra{m}\ket{\psi}^2 \bra{\psi}\ket{n}^2} \abs{\bra{m}\ket{\psi}}^2\abs{\bra{n}\ket{\psi}}^2   - \qty(\ev{O}{\psi})^2 \\
 =& {\sum_{m,n}}' \mel{m}{O}{n}^2 - \qty(\ev{O}{\psi})^2,
\end{align*}
where we use the assumption that $\braket{n}{\psi}$ and $\braket{m}{\psi}$ are real.
When we write a subspace 
in which the state $\ket{\psi}$ lives 
as $\mc{M}$, i.e., $\mc{M} = \mr{span}\qty(\{ \ket{n} | \braket{n}{\psi} \neq 0\})$,
the first term reads
\begin{align*}
 {\sum_{m,n}}' \mel{m}{O}{n}^2 &= {\sum_{m,n}}' \mel{m}{O}{n} \! \mel{n}{O}{m} \\
&= {\sum_{m}}' \bra{m}O P_{\mc{M}} O \ket{m} \\
&= \Tr_\mc{M}[ (O|_{\mc{M}})^2 ],
\end{align*}
where $P_\mc{M}$ is a projection operator onto $\mc{M}$ and $O|_\mc{M} = P_\mc{M} O P_\mc{M}$ is the operator $\mc{O}$ projected on $\mc{M}$.
Since the dimension of $\mc{M}$ grows exponentially with $N$ in general, $\Tr_\mc{M}[ (O|_{\mc{M}})^2 ]$ is $\mc{O}(2^N)$.
By considering the fact that $\ev{O}{\psi}$ is typically $\mc{O}(N)$ especially in applications to condensed matter physics and quantum chemistry, the variance of the explicit importance sampling can be exponentially large in some cases, regardless if the state is concentrated.
We remark that we here assume $f(m,n)$ can be exactly evaluated but this oversimplifies the case;
in actual applications, values of $f(m,n)$ should be also estimated by sampling and hence fluctuate for finite numbers of sampling.
We also stress that this analysis would not deteriorate the validity of our results in large quantum systems because our algorithm described in Sec.~\ref{subsec:our-method} is formally different from the explicit importance sampling.

\bibliography{bib}

\end{document}